\documentclass[reprint,amssymb,aps]{revtex4-2}
\usepackage{physics}
\usepackage{amsmath}
\usepackage{mathtools}
\usepackage{bigints}
\usepackage{nicematrix}
\usepackage{xcolor}
\usepackage{lipsum}
\usepackage{textcomp}
\usepackage{braket}
\usepackage{amssymb}
\usepackage{textcomp}
\usepackage{graphicx}
\usepackage{dcolumn}
\usepackage{bm}
\usepackage{float}

\begin{document}

\preprint{APS/123-QED}

\title{Predicting edge-localized monovacancy defects in zigzag graphene nanoribbons from Floquet quasienergy spectrum }

\author{Gulshan Kumar}
\author{Shashikant Kumar}
\author{Ajay Kumar}
\author{Prakash Parida}
 \email{pparida@iitp.ac.in}
\affiliation{%
  Department of Physics,
  Indian Institute of Technology Patna,\\
  Bihta, Patna, Bihar 801106, India\\
  }

\begin{abstract}
In this work, we prescribe a theoretical framework aiming at predicting the position of monovacancy defects at the edges of zigzag graphene nanoribbons (ZGNRs) using Floquet-Bloch formalism, which can be experimentally observed through time- and angle-resolved photoemission spectroscopy (tr-ARPES). Our methodology involves an in-depth investigation of the Floquet quasienergy band spectrum influenced by light with varying polarization across a range of frequencies. Particularly under the influence of circularly polarized light with a frequency comparable to the bandwidth of the system, our findings suggest a promising approach for locating monovacancy defects at either edge, a challenge that proves intricate to predict from the ARPES spectrum of ZGNRs with monovacancy defects. This has been achieved by analyzing the orientation of the Floquet edge state and the appearance of new Dirac points in the vicinity of the Fermi level. The real-world applications of these captivating characteristics underscore the importance and pertinence of our theoretical framework, paving the way for additional exploration and practical use. Our approach, employing the Floquet formalism, is not limited to monovacancy-type defects; rather, it can be expanded to encompass various types of vacancy defects.

\end{abstract}
\maketitle
\section{Introduction}
 Graphene, known for its structural flexibility and exceptional strength, can be transformed into a diverse array of derivative structures, including carbon nanotubes, fullerenes, quantum dots and ribbons, each possessing remarkable physical and electrical attributes \cite{ref1,ref6}. However, among these derivative structures, graphene nanoribbons (GNRs), specifically the two types known as zigzag graphene nanoribbons (ZGNRs) and armchair graphene nanoribbons (AGNRs), receive the most extensive research attention \cite{ref7,ref8,ref9,ref10,ref11}.  

The study was expanded to encompass GNRs featuring various types of defects, which exert an influence on their physical characteristics. These defects specifically impact the transport properties \cite{defe_transport}, electronic properties \cite{defe_conduct}, mechanical properties \cite{defe_mech}, and magnetism \cite{defe_magn} of the GNRs under investigation. During the fabrication of graphene nanoribbons, vacancy defects may arise naturally, but they can also be intentionally introduced or controlled to suit specific objectives in nanoribbon design and functionality  \cite{ref12,ref13,ref14,ref15,ref16,ref17,ref18,ref19}. Various methodologies can be utilized for this purpose, encompassing top-down techniques such as focused ion beams or electron beam lithography \cite{ref20,ref21,ref22}, along with chemical processes like oxygen plasma treatment or chemical vapor deposition, to intentionally introduce vacancy defects into zigzag graphene nanoribbons (ZGNRs) \cite{ref23,ref24}. Moreover, Schmidt et al. emphasized helium ion beam milling as a promising technique for creating smaller pores with reduced inter-defect distances. Their successful fabrication of $3 nm$ to $4 nm$ pores with a pitch of $10 nm$ showcased the effectiveness of this method \cite{schmidt}. Employing lithography techniques, a tailored nanoscale superstructure of variable design and dimensions, reaching down to tens of nanometers, can be fabricated. Recently, Jones et al. conducted experimental work showcasing the controlled periodic patterning of nanoscale pores in graphene \cite{n_arpes1}. They demonstrated precise control over the induction of massive Dirac fermions by manipulating the size of nanoscale apertures in graphene. In their density functional theory (DFT) investigation, they focused on a pore size of $0.7 nm$. However, top-down approaches encounter challenges in achieving precise periodicity for a few atomic vacancy defects in crystal structures with the current experimental setups available. 
	
To bridge this limitation, one can utilize bottom-up methodologies, including on-surface and in-solution techniques, to craft atomically precise Zigzag Graphene Nanoribbons (ZGNRs) by employing tailored precursor monomers \cite{cape3}. Similar approaches have been employed in synthesizing edge-extended ZGNRs and creating graphene nanoribbon heterojunctions and heterostructures \cite{motifs1,motifs2}, particularly focusing on ZGNRs with modified edges featuring periodic cove-cape units \cite{cape1}. Liu et al. successfully demonstrated the precise formation of cove-type edges in ZGNRs both on substrates and in solution, utilizing chrysene as a crucial monomer \cite{cove2}. Wang et al. illustrated the tunability of the bandgap in cove-edge ZGNRs through a balance between the length of the zigzag segment and the distance between adjacent cove units \cite{cove1}. Notably, they highlighted that graphene nanoribbons with cove edges can be viewed as periodically removing a carbon atom at the edge of corresponding fully formed ZGNRs. The availability of experimentally developed periodic structures has inspired us to select our system, which features monovacancy defects at the edges of ZGNRs.

To detect different defects in 2D-structured materials, there are various experimental techniques, such as transmission electron microscopy (TEM) and scanning tunneling microscopy (STM) \cite{ref28,ref29,ref30,ref31,ref32,ref33,ref34}. J. C. Mayers et al. employ TEM to detect single vacancy defects and edge defects \cite{ref29}, whereas H. Yan et al. utilize TEM and STM imagery to investigate structural anomalies \cite{ref33}. Furthermore, AFM and Raman spectroscopy find application in the examination of the surface morphology of defective graphene \cite{ref35,ref36}. 

It has been demonstrated that the geometric structure and electronic properties are significantly influenced by the type and location of defects \cite{ref26,ref27}. The size of the defect also plays an important role in the band structure, as demonstrated by N. Nourie et al. and T.G. Pedersen et al., who theoretically showed the size effect of antidots in silicene \cite{silicene} and graphene \cite{ref12}, respectively. With an increase in the size of antidots, the band gap will increase at the Fermi level. A. J. Jones et al. verified this experimentally using angle-resolved photoemission spectroscopy with nanoscale spatial resolution (n-ARPES) \cite{n_arpes1} for different diameters of vacancies in 2D graphene. They have studied the effect of nanoscale holes in the graphene sheet by measuring effective mass of carriers and band gaps using n-ARPES. It is very hard to observe the effect of nano holes in the n-ARPES spectrum. However, electrostatic doping enhances the effective mass and leads to the direct observation of an electronic bandgap in their n-ARPES spectrum. Thus, with(without) doping, it is really challenging to observe any effect of a hole of very small size (for example, a single-atom vacancy defect)  in the ARPES spectrum of graphene. 

To overcome the challenge of observing the effect and, hence, the presence of monovacancy in the ARPES spectrum, we introduce the an application of the Floquet theory to identify the locations of monovacancy defects situated at the edges of zigzag graphene nanoribbons from the Floquet quasienergy spectrum. We note that, Floquet quasienergy spectra have already been observed experimentally using time- and angle-resolved photoemission spectroscopy (tr-ARPES) for a few other systems \cite{trARPES1,trARPES2,trARPES3}. Additionally, for the 1D ribbon, edges play a crucial role in the formation of the band structure. Therefore, understanding the position of vacancy defects within the crystal is crucial for comprehending the properties of the system. We investigate the monovacancy defects situated at the edges of ZGNRs, while considering the influence of light with different polarization and adjustable parameters. Building on our observations, our goal is to develop a methodology that predicts the location of a monovacancy defect at a specific edge of a ZGNR. This will be achieved through an analysis of the interaction between circularly polarized light and ZGNRs featuring a monovacancy defect at the edge. 

\section{\label{sec:levelI}FLOQUET-BLOCH THEORY}
When a ZGNR, periodic in x-direction with lattice vector ${\bf b} = (\sqrt{3}a_0l,0)$ and nearest neighbor (NN) vector $\delta_j$ given by $\delta_1 = a_0(0,1)$, $\delta_2 = a_0/2(-\sqrt{3},-1)$ and $\delta_3 = a_0/2(\sqrt{3},-1)$ where `$a_0$' is the bond length and `$l$' is the count of unitcell in a supercell, is exposed to an electromagnetic field characterized by the vector potential ${\bf A}(\tau) = (A_x \cos(\omega \tau), A_y \cos(\omega \tau+\phi))$ with arbitrary amplitude and phase, the interaction between light and electrons can be effectively described using the Peierls substitution, ${\bf k} \rightarrow {\bf k} + \frac{e{\bf A}(\tau)}{c}$ \cite{peierl, peierl_2}. This approach captures the spacial and temporal periodicity exhibited by the resulting quantum system, as reflected in the time-dependent tight-binding Hamiltonian ${\bf H}({\bf x},\tau)$ governing its behavior. The periodicity is characterized by the condition ${\bf H}({\bf x + {\bf b}},\tau) = {\bf H}({\bf x},\tau+T)$, where $T = 2\pi/\omega$ represents the time period. To analyze such systems, the Floquet technique is employed, which allows for the study of time-periodic Hamiltonian. We can assume eigenstates of the time-periodic Hamiltonian in the Floquet-Bloch form $\ket{\Psi_{\alpha,\vb{k}}{(\vb{x},\tau)}} = e^{i\vb{k}.\vb{x}-i\varepsilon_{\alpha,\vb{k}}\tau/\hbar}\ket{u_{\alpha,\vb{k}}{(\vb{x},\tau)}}$, where $\varepsilon_{\alpha,\vb{k}}$ represents the quasienergy and $\ket{u_{\alpha,\vb{k}}{(\vb{x},\tau)}}$  represents the $\alpha$  Floquet state with {\bf k} wave vector. Due to the time translational invariance of system, the time-dependent Schr\"{o}dinger equation is transformed into an eigenvalue equation, ${\bf H}_F(\vb{k},\tau)\ket{u_{\alpha,\vb{k}}{(\tau)}} = \varepsilon_{\alpha,\vb{k}}\ket{u_{\alpha,\vb{k}}{(\tau)}}$, with ${\bf H}_F({\bf k},\tau) = {\bf H}_{\bf k}(\tau) - i\hbar\pdv{}{\tau}$ as the Floquet operator and ${\bf H}_{\bf k}(\tau)$ is the time dependent Bloch Hamiltonian. As we know that Floquet state is periodic in time, we can expand it in terms
of Fourier components $\ket{u_{\alpha,\vb{k},p}}$.

\begin{equation}
	\ket{u_{\alpha,\vb{k}}{(\vb{x},\tau)}} = \sum_{p}e^{-i p\omega\tau}
	\ket{u_{\alpha,\vb{k},p}}  
\end{equation}

To deal with basis {$\ket{u_{\alpha,\vb{k},p=1,2,3..}}$}, one has to consider
composed Hilbert space $\mathcal{S}=\mathcal{H}\otimes\mathcal{T}$, where
$\mathcal{H}$ is the Hilbert space and $\mathcal{T}$ is time-periodic space \cite{sambe}.
Quasienergies can be obtained by the use of composed scalar product which
allows the diagonalisation of matrix,

\begin{equation}
\begin{split}
    \bra{u_{\alpha,\vb{k}}(\tau)}{\bf H}_F(\vb{k},\tau)
	\ket{u_{\alpha,\vb{k}}{(\tau)}} =  \\     
	\sum_{p,q}\bra{u_{\alpha,\vb{k},p}}{\bf H}_{p-q}(\vb{k})
	\ket{u_{\alpha,\vb{k},q}}-\delta_{p,q}q\hbar\omega 
\end{split}
\label{FEVE}
\end{equation}

Where Fourier components of Floquet Hamiltonian ${\bf H}_{p}(\vb{k})$ are given by Eq. \ref{integral_equation}.
\begin{equation}\label{integral_equation}
	{\bf H}_{p}(\vb{k})=\frac{1}{T}\int^T_0 d\tau
	 {\bf H}_{\vb k}(\tau)e^{ip\omega\tau}
\end{equation}
Furthermore, the infinite-dimensional Floquet Hamiltonian can be expressed in matrix form within the basis of the composed Hilbert space. 
\[
 {\bf H}_F = \begin{bmatrix} 
    \ddots & \vdots & \vdots & \vdots & \vdots & \iddots \\
    \dots & H_0-2\hbar\omega &  H_1  &  H_2  & H_3 & \dots \\
    \dots &  H_{-1} & H_0-\hbar\omega & H_1 & H_2 & \dots \\
    \dots &  H_{-2} & H_{-1} & H_0 & H_1  & \dots \\
    \dots &  H_{-3} & H_{-2} & H_{-1} &H_0+\hbar\omega& \dots \\
    \iddots & \vdots  & \vdots & \vdots & \vdots & \ddots \\
    \label{Floquet_matrix}
    \end{bmatrix}
\]
Matrix ${\bf H}_F$ is organized in a way that its diagonal blocks ($H_{m=0}$) represent distinct Floquet sidebands, each with an energy separation of $\hbar\omega$. Conversely, the off-diagonal
blocks ($H_{m=\pm 1,\pm 2,\pm 3...}$) capture interactions among these sidebands, which originate from the impact of time-dependent fields. 
The size of these matrix blocks is contingent upon the basis of the undriven ZGNR. In these off-diagonal blocks, the strength of coupling between adjacent sites is contingent upon the vector potential ${\bf A}(\tau)$. The renormalised hopping integrals are ${t}_{j,m}^F = te^{im{\bf \xi}_j}J_m(\eta_j)$, where index $j = 1,2,3 $ is for the nearest neighbour vector ($\delta_j$) and $J_m$ represents the $m^{th}$ order Bessel function. All the information of field configuration are encoded in the dimensionless quantity $\eta_j$ and $\xi_j$. \\
\begin{align*}
	\centering
\eta_{2,3} &= \frac{ea}{2c}\sqrt{3A^{2}_{x}+A^{2}_{y}
        \pm 2\sqrt{3}A_{x}A_{y}\cos{\phi}} \\     
\xi_{2,3} &=\mp{\left(\frac{\pi}{2}+\tan^{-1}{\left(\frac{\sqrt{3}A_{y}\sin{\phi}}{A_{x}
        \pm \sqrt{3}A_{y}\cos{\phi}}\right)}\right)} \\
\eta_{1} &= A_xea/c,\hspace{2cm}  \xi_1 = \pi/2-\phi
\end{align*}
Hereafter physcial parameters llike $\hbar$, $e$ and $c$ is set as $1$ to make the field parameter dimensionless. To do numerical computation, we employ the strategy of photon-number truncation method to effectively reduce the infinite-dimensional Floquet-Bloch matrix (${\bf H}_F$) into a reduced-size Floquet-Bloch matrix, which exclusively encompasses a finite set of Fourier components derived from the Hamiltonian. In our scenario, we implement a Hamiltonian cutoff based on a two-photon number criterion. This entails considering the Fourier component ${\bf H}_{\pm m}$, where the parameter $m$ ranges from zero to two. [for further explanation see Appendix (a)]

\section{RESULTS AND DISCUSSIONS}

\begin{figure}[ht]
 	\includegraphics[width=8.5cm,height=4.5cm]{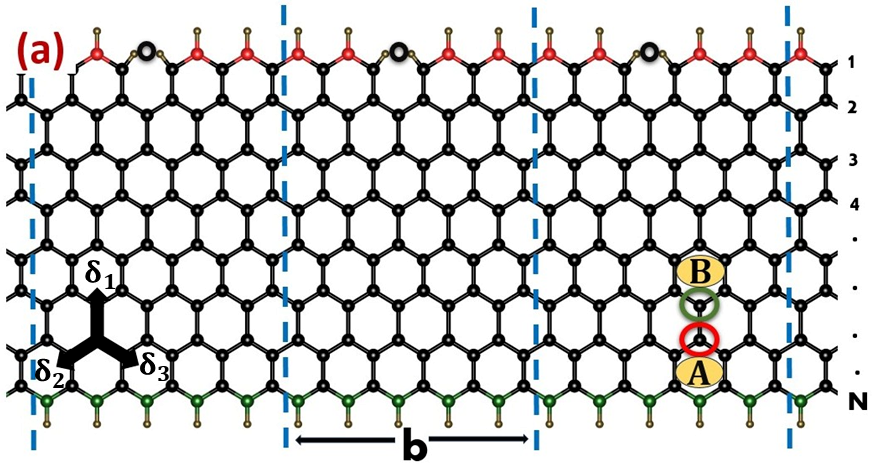}    
	\includegraphics[width=8.5cm,height=4.5cm]{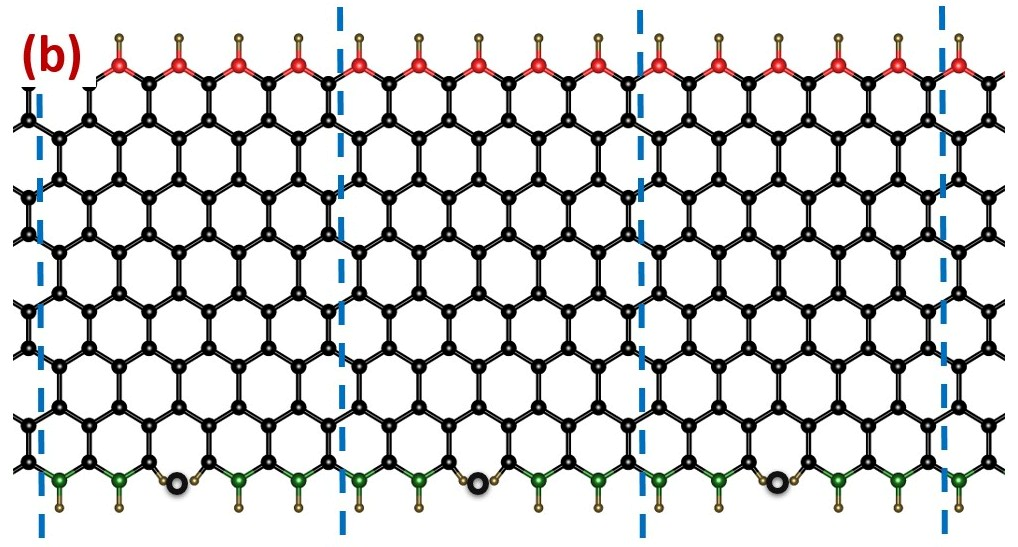}    
	\caption{Demonstration of ZGNRs with array of monovacancy defects at edge of (a) lattice type-A represented by red-colored atoms and (b) lattice type-B represented by green-colored atoms. The monovacancy site is represented by a blue circle at the edges. $\delta_1$, $\delta_2$, and $\delta_3$ represent the nearest neighbor vectors. ${\bf b}$ is the lattice vector of the supercell, represented by dashed lines.}
	\label{zgnr_monovacancy_image}
\end{figure}

In our study, we investigate monovacancies positioned at the edges of ZGNRs. These monovacancies result from the removal of a carbon atom from the hexagonal lattice, creating a single vacant site defect. Unlike bulk graphene, where vacancies exhibit Jahn-Teller distortion, the edge monovacancies in ZGNRs have unique properties \cite{jahntellar1,jahntellar2}. Monovacancies at the ZGNR edges create two dangling bonds, which are effectively passivated by hydrogen atoms, thus stabilizing the edge structure. We perform density functional theory (DFT) calculations to verify the existence of monovacancies at the edge using the VASP package. We choose two initial structures, each sharing the same unit cell but differing in the arrangement at the edge: one featuring a pentagon motif with no dangling bonds, and the other displaying a monovacancy where two dangling bonds are saturated with hydrogen atoms. Both structures undergo optimization across all degrees of freedom, including atomic relaxation and lattice constants, until the force on each atom reaches $0.01 eV/\AA$. We calculate formation energies per unit cell for both the optimized pentagon motif (referred to as ``PT'') and the monovacancy (referred to as ``MV'') structures. The computational results reveal that the monovacancy at the edge exhibits greater energetic stability compared to the formation of a pentagon motif. Specifically, the formation energy of the monovacancy is determined to be $-1.13 eV$, while for the pentagon motif, it is $2.77 eV$. Additionally, we calculate the energy change associated with a reaction where the initial state is the MV structure transitioning to the formation of a pentagon motif structure while liberating a hydrogen ($H_2$) molecule as a product (MV structure $\rightarrow$ PT structure + $H_2$). The computed reaction energy, $\Delta E = 3.96 eV$, indicates an endothermic process, implying that energy input is required for converting the MV structure into the pentagon motif while releasing the hydrogen molecule. This observation aligns with findings reported by Sahan et al. \cite{DFTmv}, who also observed a similar energy difference between monovacancy and pentagon motifs at the edge of ZZNRs. Furthermore, the bond length between two dangling bonded atoms for forming the pentagon was found to be approximately 1.89 Å, which closely matches their reported value of 1.8 Å. The computational findings suggest that a monovacancy is energetically more favorable than a pentagon motif, indicating that optimizing the edge structure would entail avoiding the formation of pentagons. This is supported by the results suggesting that the release of $ H_2 $ is less probable, and the defected edge behaves similarly to a notched hexagonal edge (For computational detail see Appendix B). These monovacancy can manifest at two distinct positions within the lattice, namely sublattice A and B, as illustrated in Fig. \ref{zgnr_monovacancy_image}(a). We note that, one edge of ZGNR terminates with sublattice type-A and the other edge terminates with sublattice type-B. Hence, two edges of ZGNRs are not equivalent.

We first delve into the band structure characteristics of hydrogen-passivated ZGNRs exhibiting monovacancy defects either at the edge of sublattice type-A (A-ZGNR), depicted by red-colored atoms in Fig. \ref{zgnr_monovacancy_image}(a), or at the edge of sublattice type-B (B-ZGNR), represented by green-colored atoms in Fig. \ref{zgnr_monovacancy_image}(b). A-ZGNR represents vacancy at one of the edge, while B-ZGNR represents vacancy at the other edge. For our investigation, we have chosen a ZGNR with a width defined by $N = 20$ carbon chains, as depicted in Fig. \ref{zgnr_monovacancy_image}(a). The movement of $\pi$-electrons within a nanoribbon featuring periodic defects can be elucidated through the utilization of the tight-binding (TB) model.

\begin{figure}[ht]
	\centering
	\includegraphics[width=4cm,height=4cm]{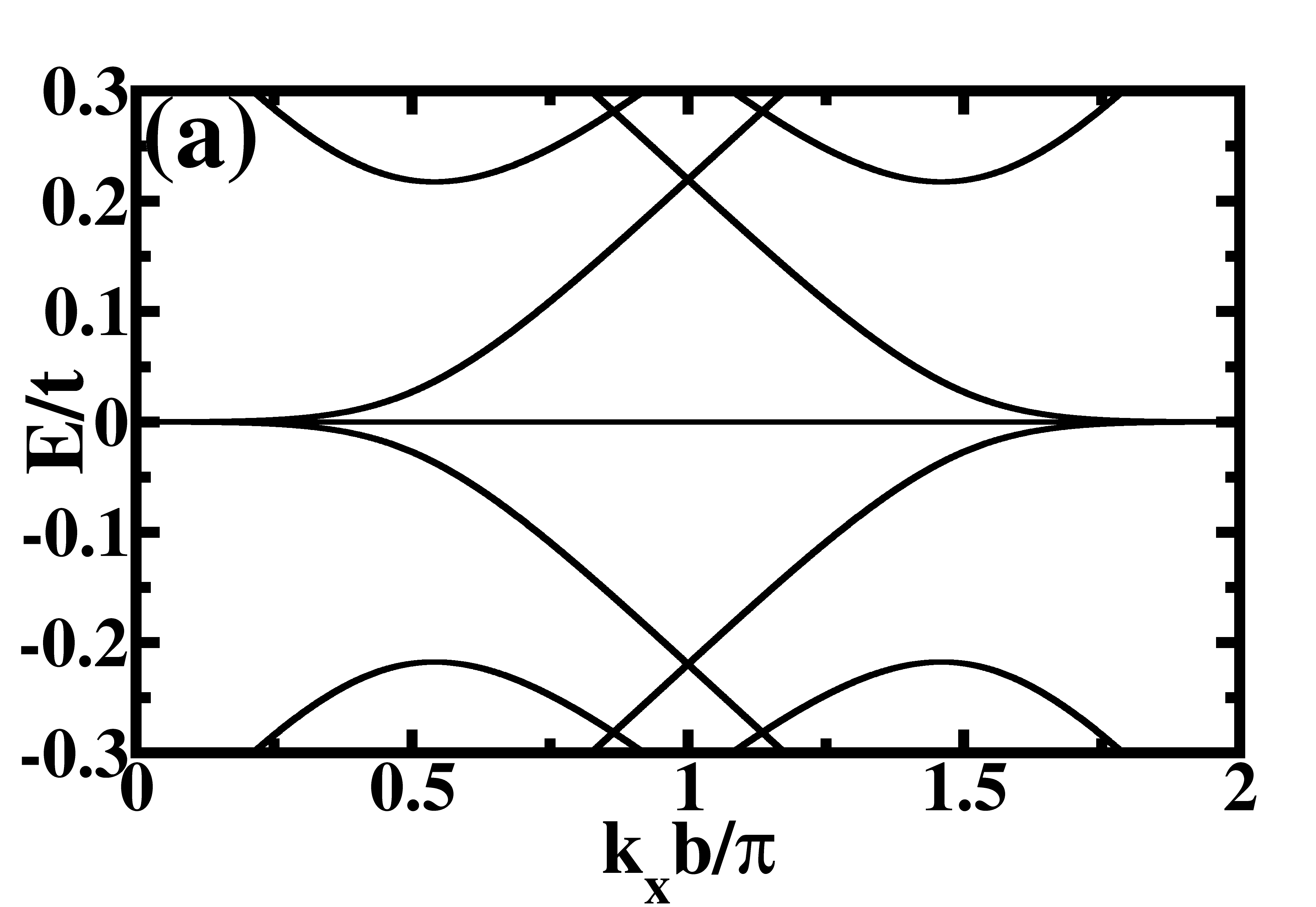}
	\includegraphics[width=4cm,height=4cm]{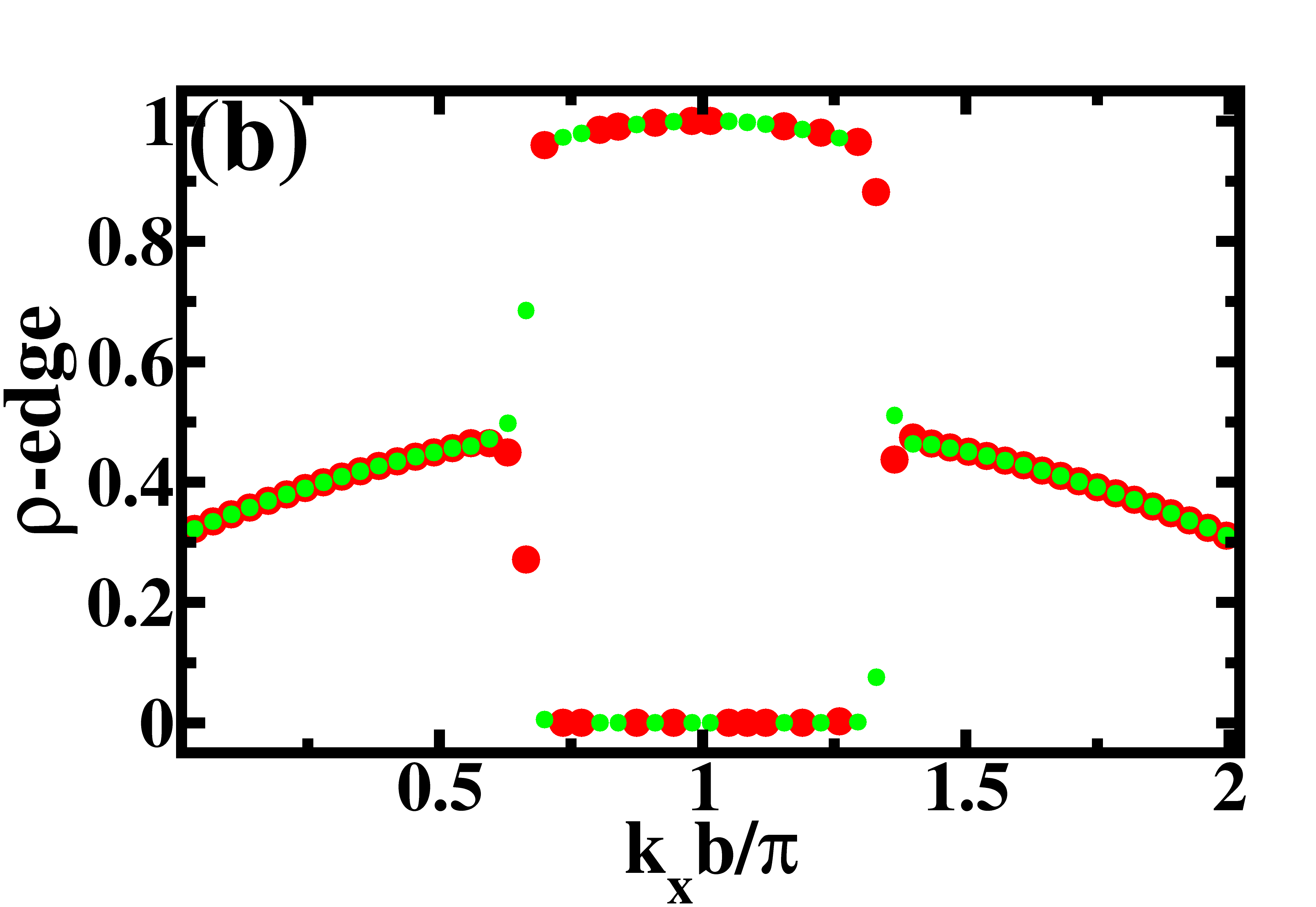}
	\includegraphics[width=4cm,height=4cm]{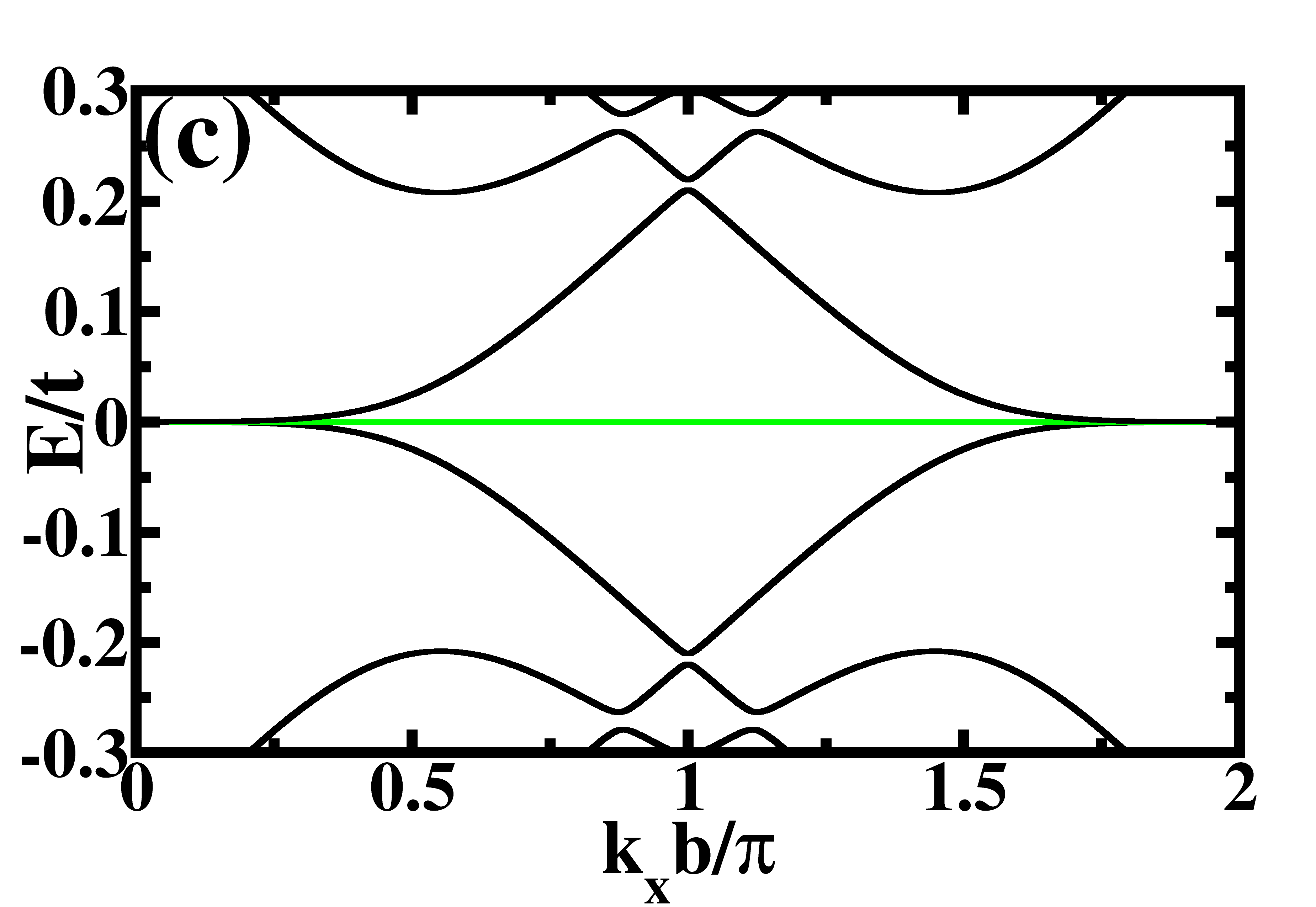}
	\includegraphics[width=4cm,height=4cm]{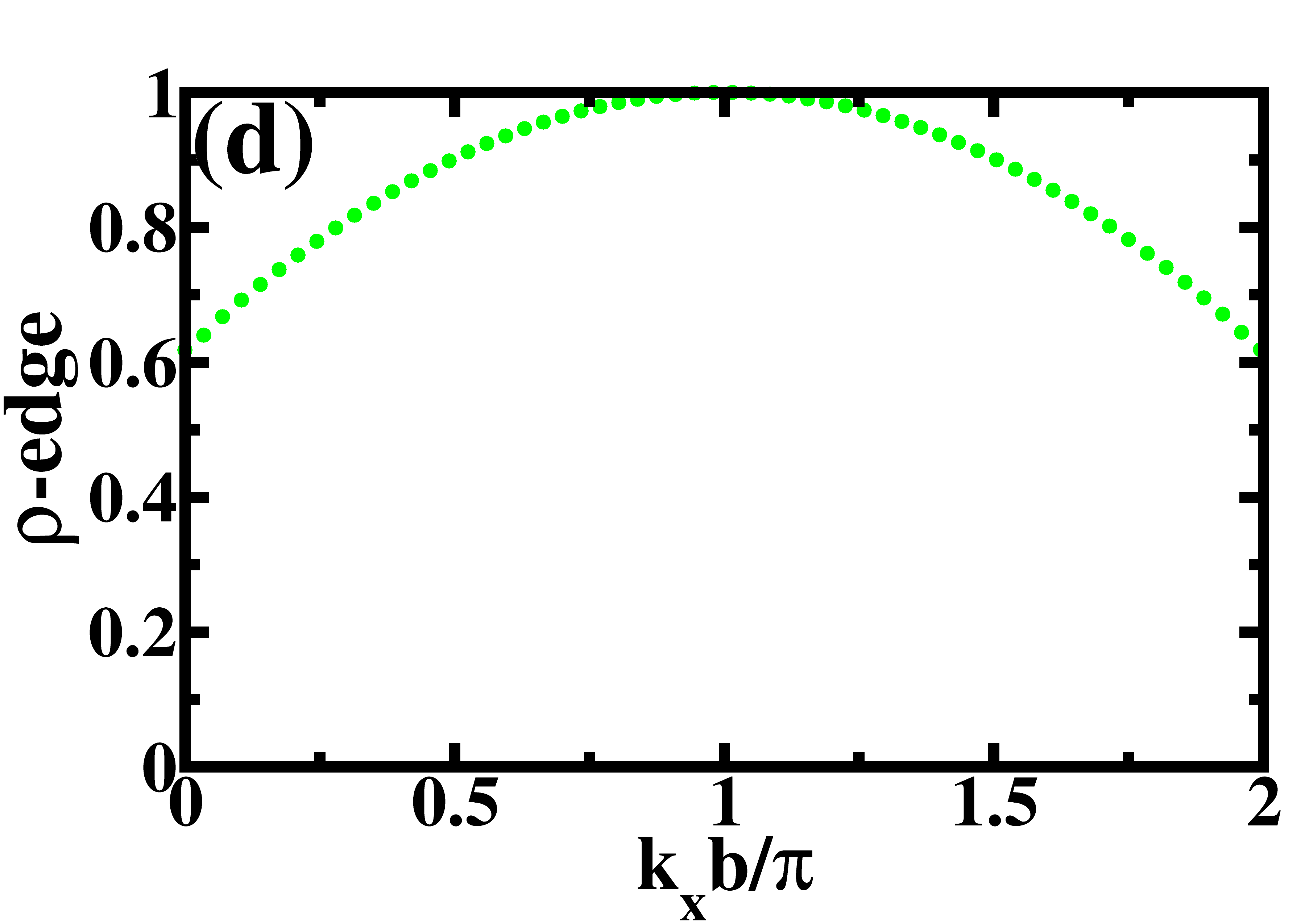}
	\includegraphics[width=4cm,height=4cm]{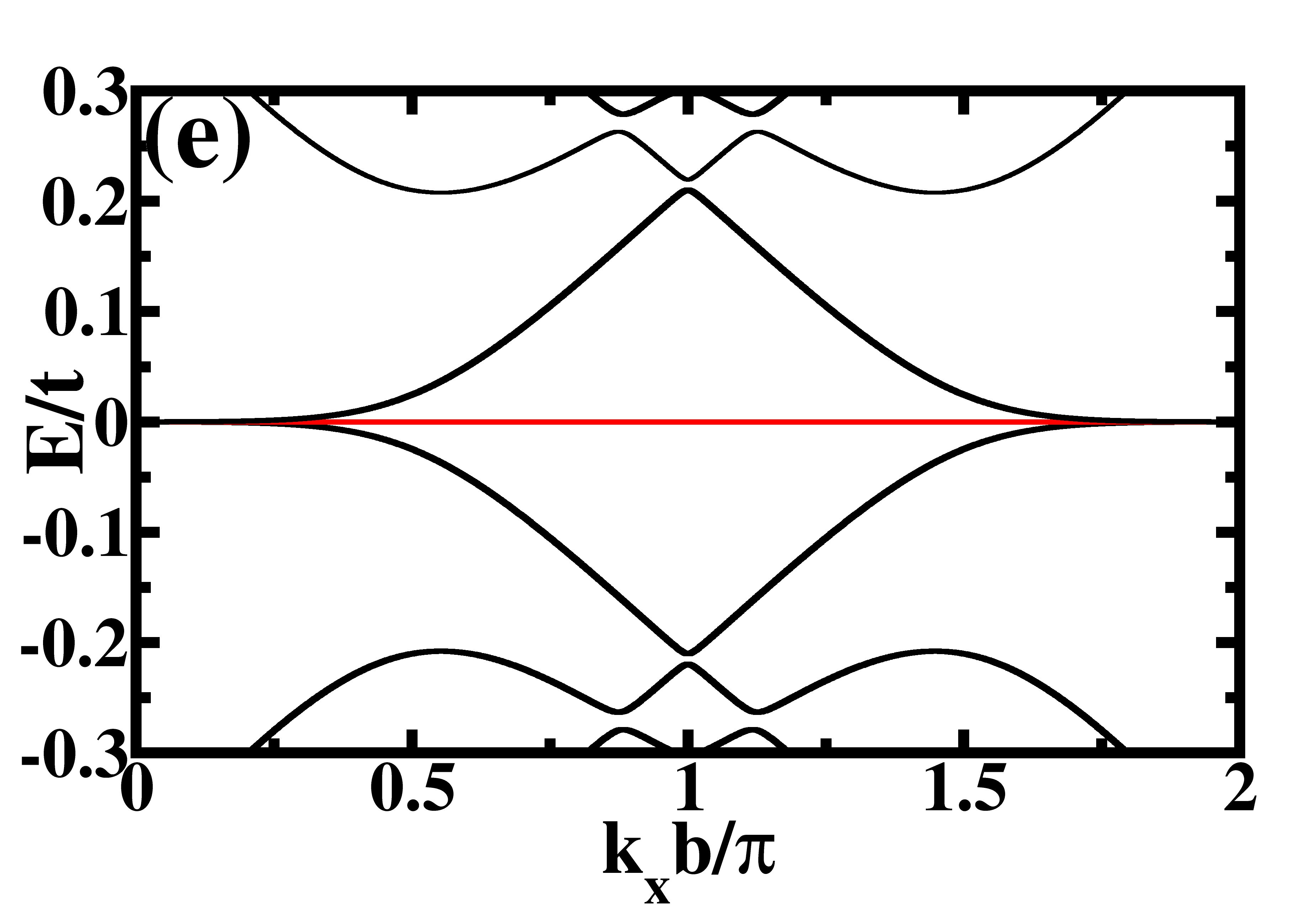}
	\includegraphics[width=4cm,height=4cm]{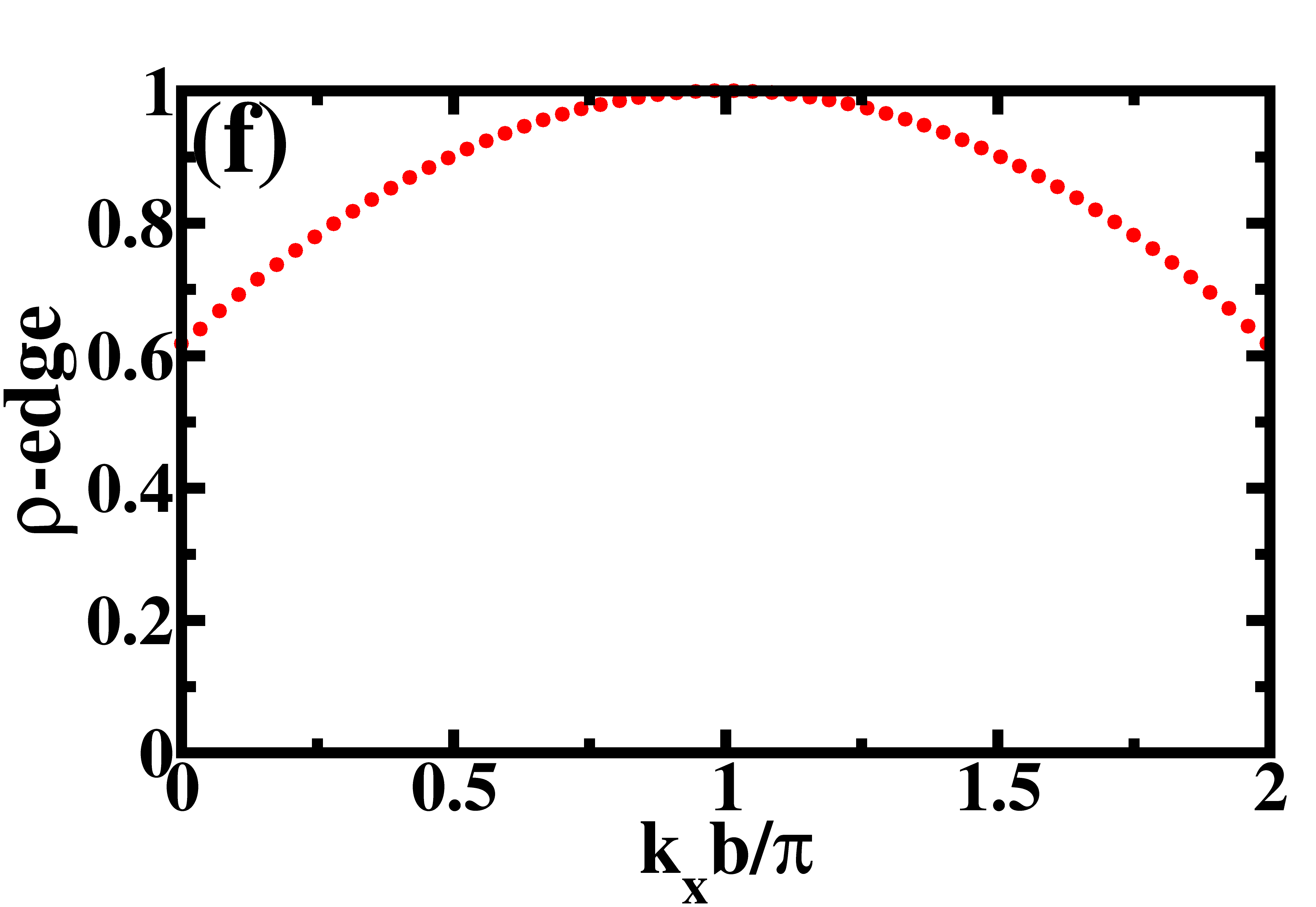}
	\caption{The plots showcase the electronic band structures of pristine ZGNR in panel (a), A-ZGNR in panel (c), and B-ZGNR in panel (e). Panel (b) details the contribution of  EDD of edge states the to one of the flat bands at zero energy in ZGNR. Similarly, panels (d) and (f) depict the contributions of EDD edge states to the flat bands at zero energy in A-ZGNR and B-ZGNR, respectively. Red color depicted to the edge of lattice type-A and green color depicted to the edge of lattice type-B.}
	\label{monovacancy_band_normal}
\end{figure}

\begin{figure*}
	\includegraphics[width=5.25cm,height=5.5cm]{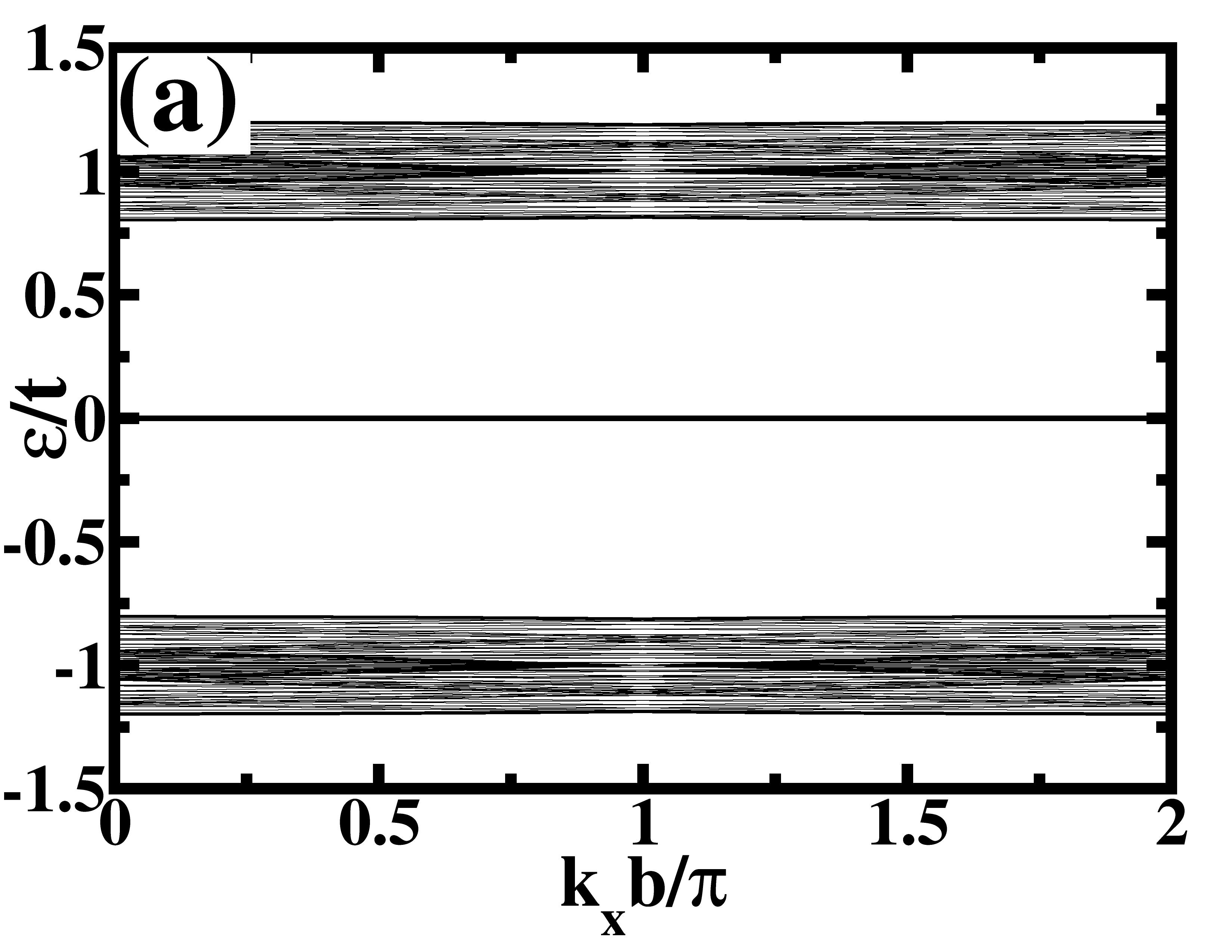}
	\includegraphics[width=5.25cm,height=5.5cm]{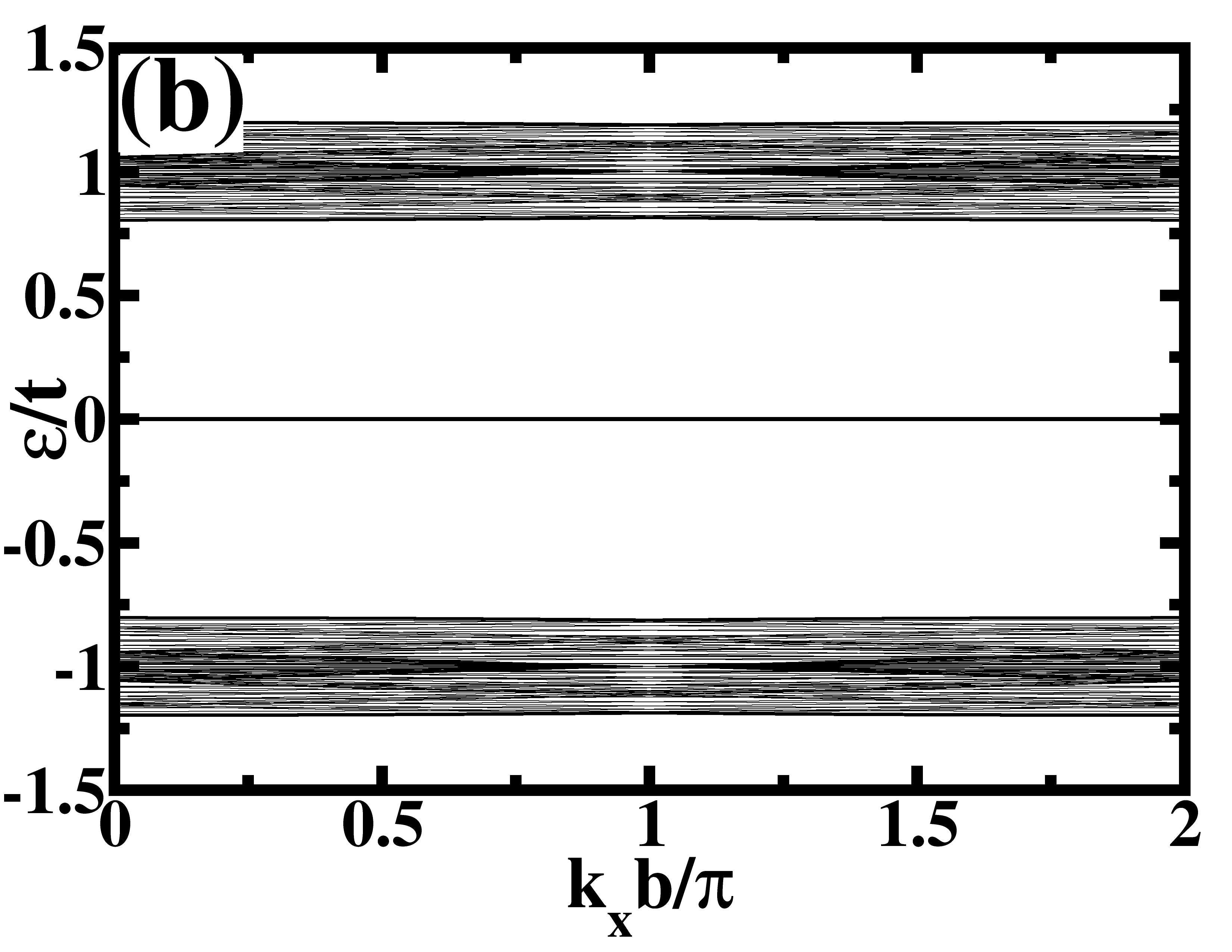}
	\includegraphics[width=5.25cm,height=5.5cm]{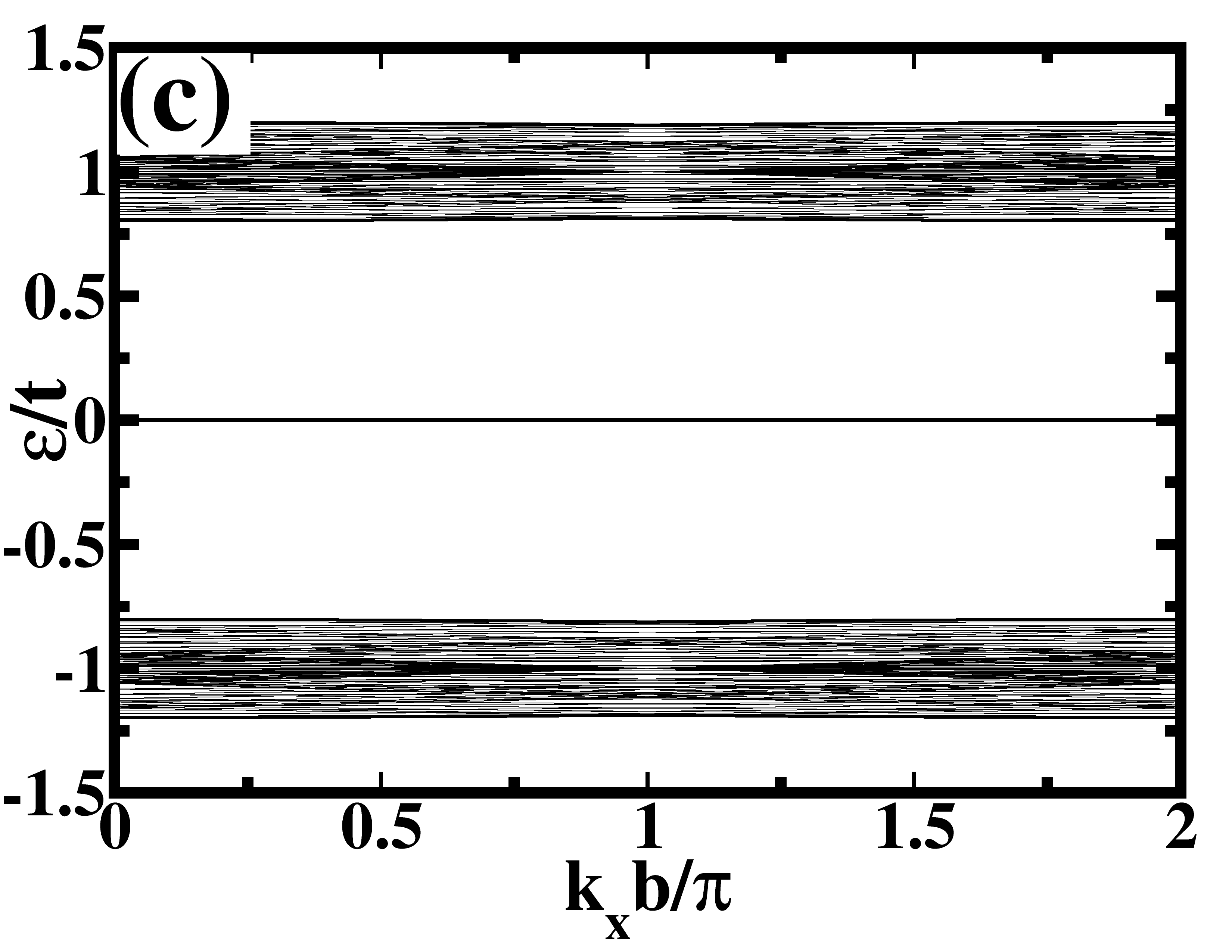}
	\includegraphics[width=5.25cm,height=5.5cm]{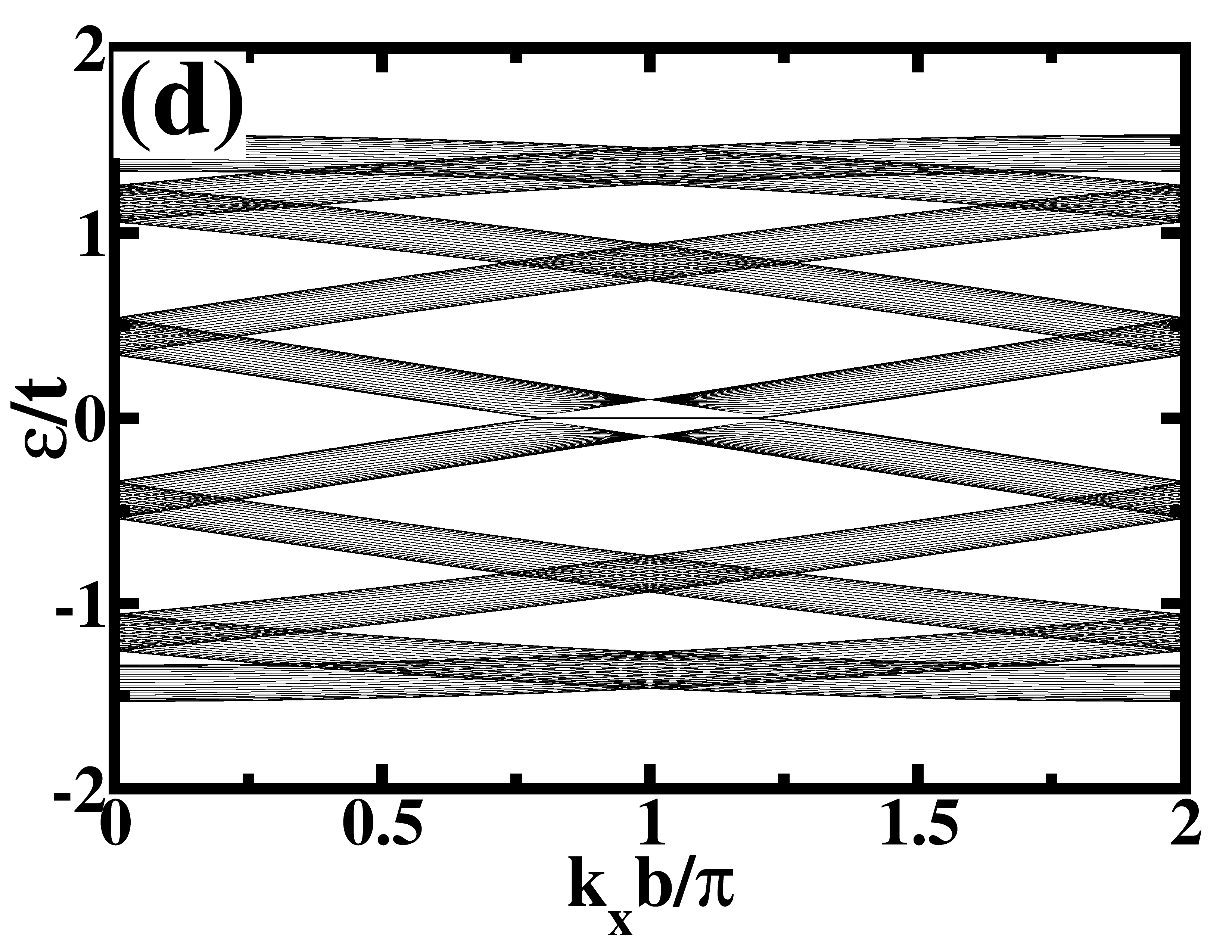}
	\includegraphics[width=5.25cm,height=5.5cm]{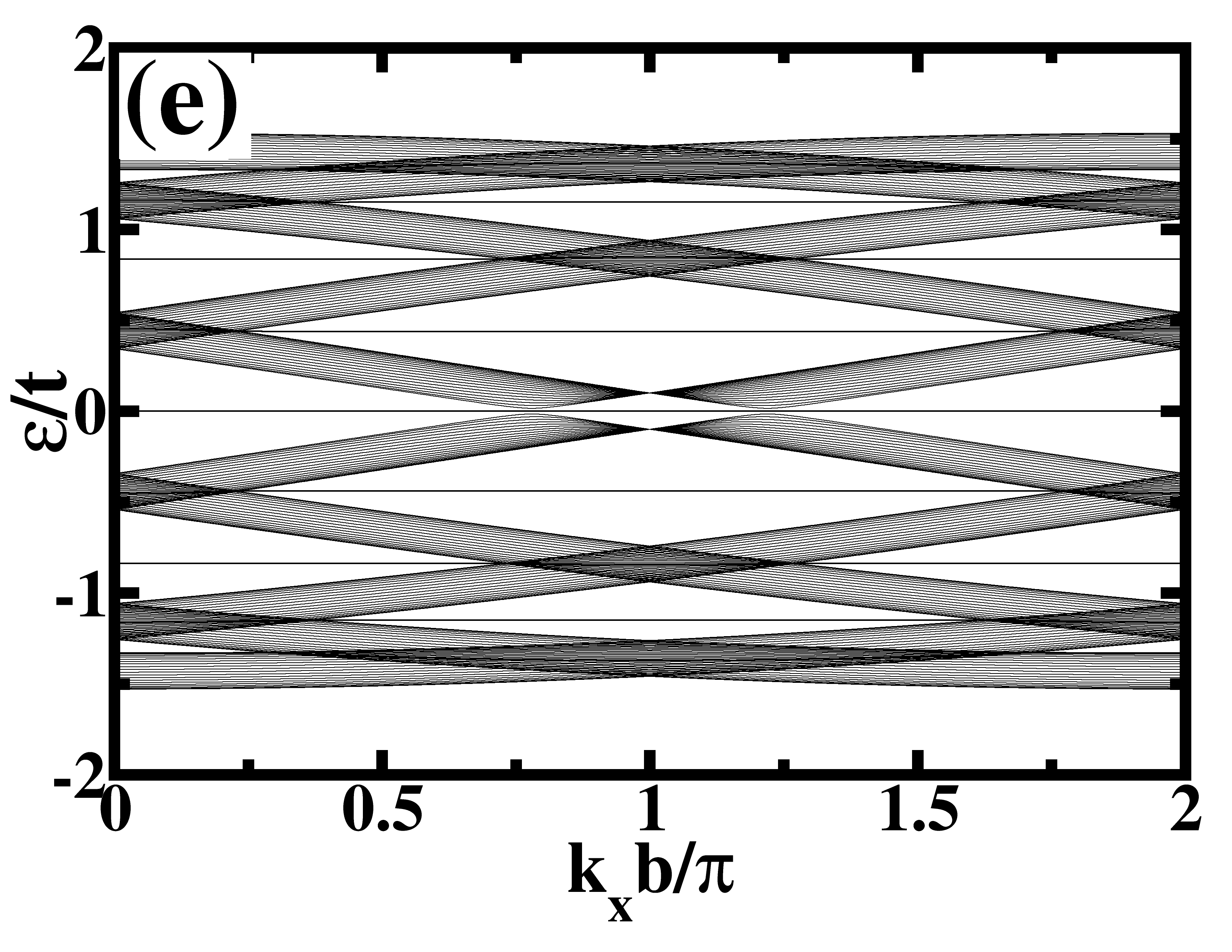}
	\includegraphics[width=5.25cm,height=5.5cm]{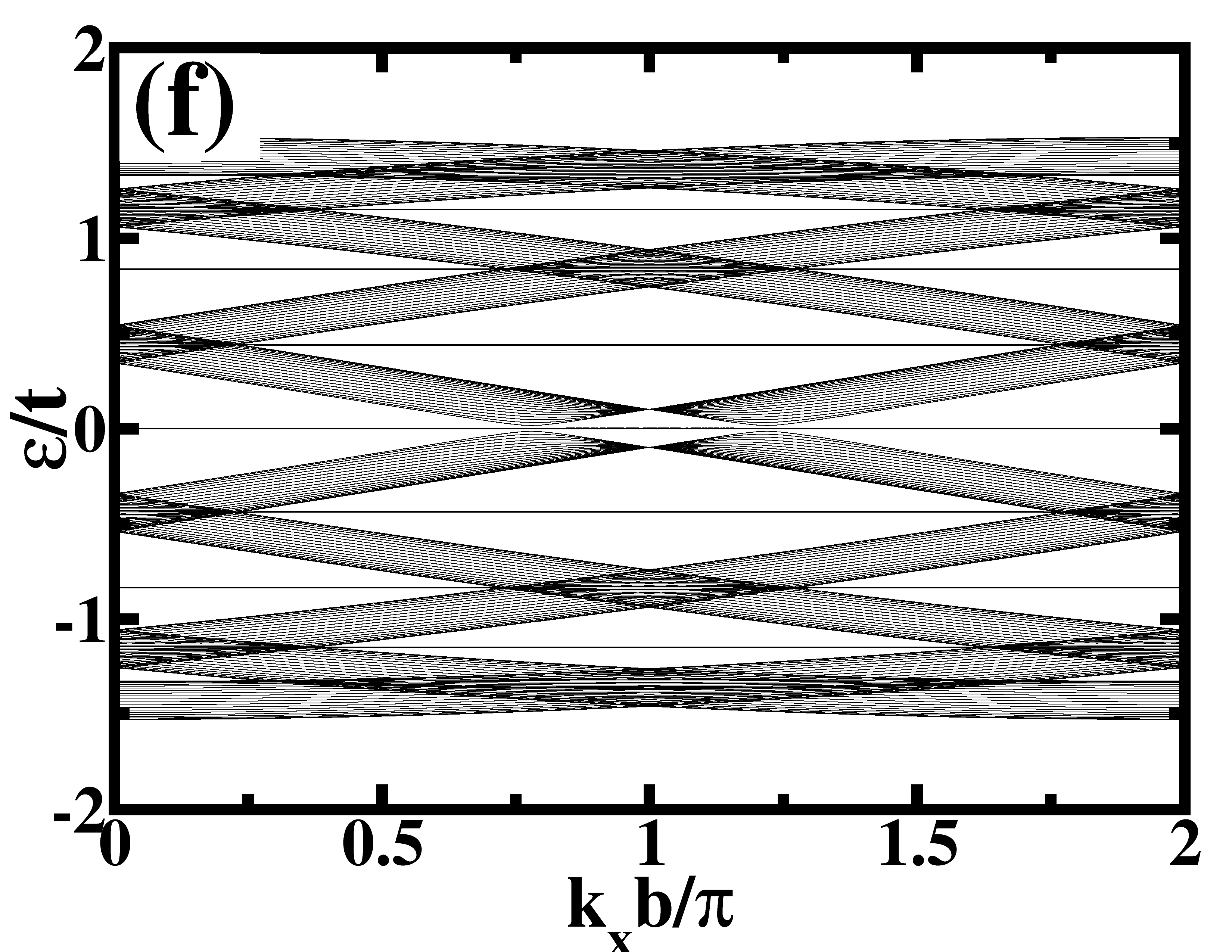}
	\includegraphics[width=5.25cm,height=5.5cm]{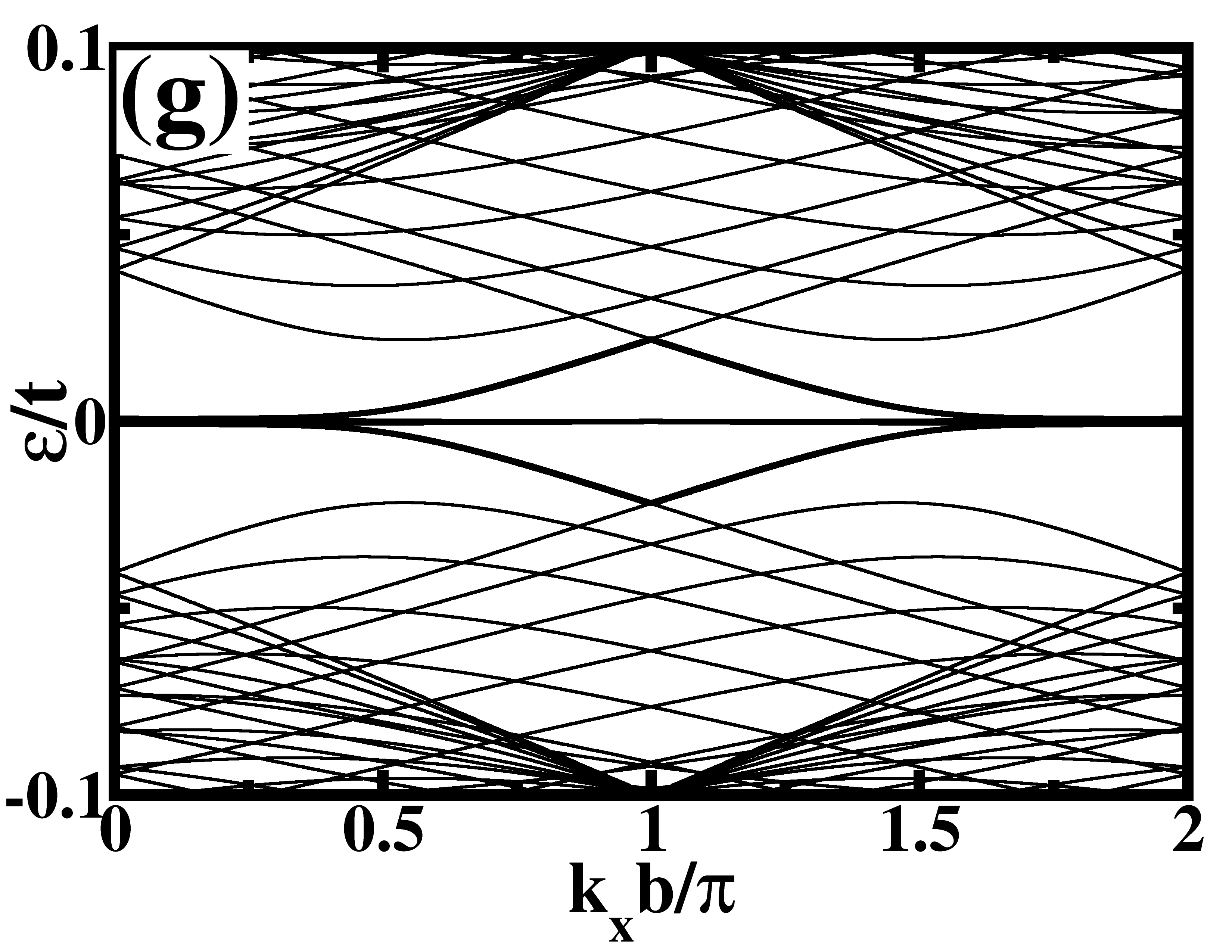}
	\includegraphics[width=5.25cm,height=5.5cm]{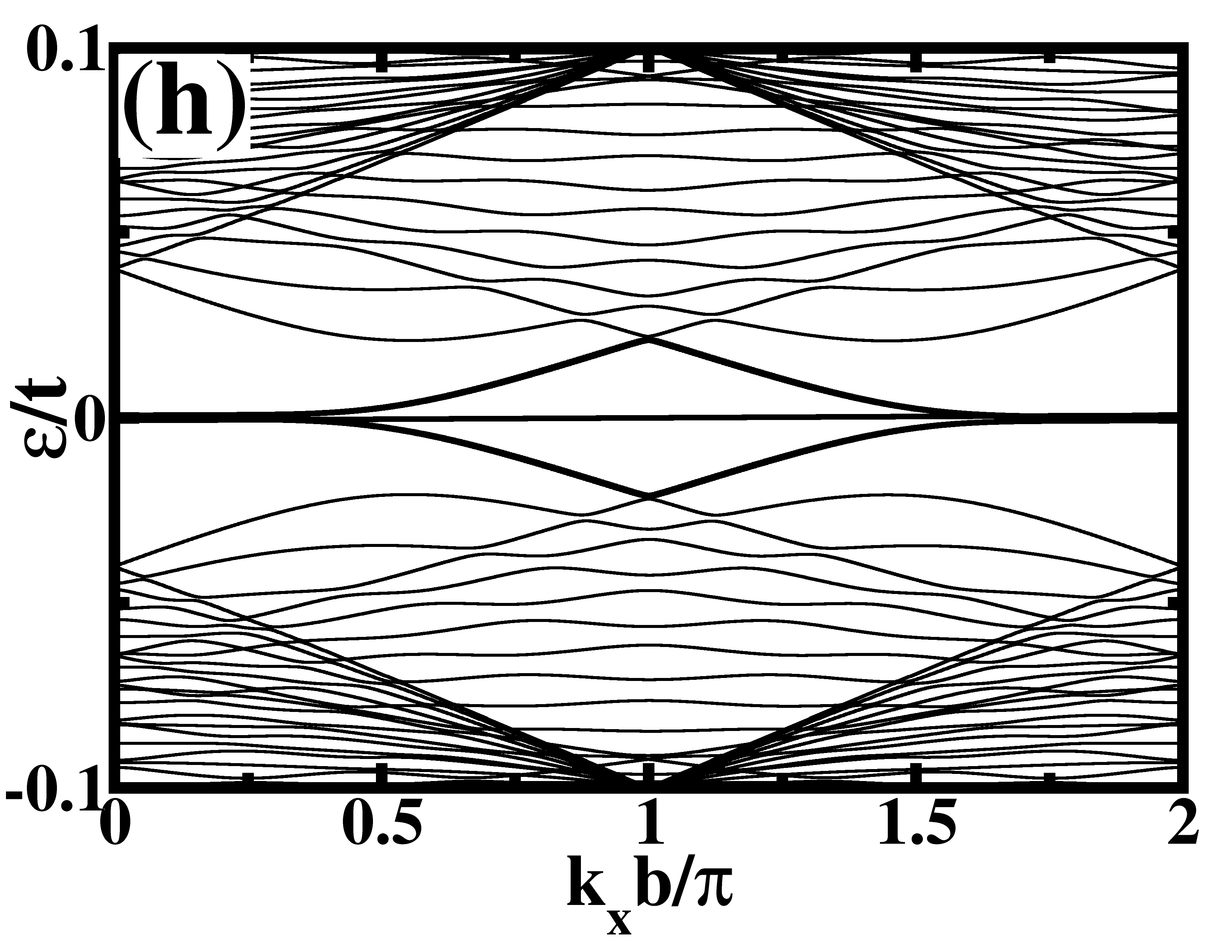}
	\includegraphics[width=5.25cm,height=5.5cm]{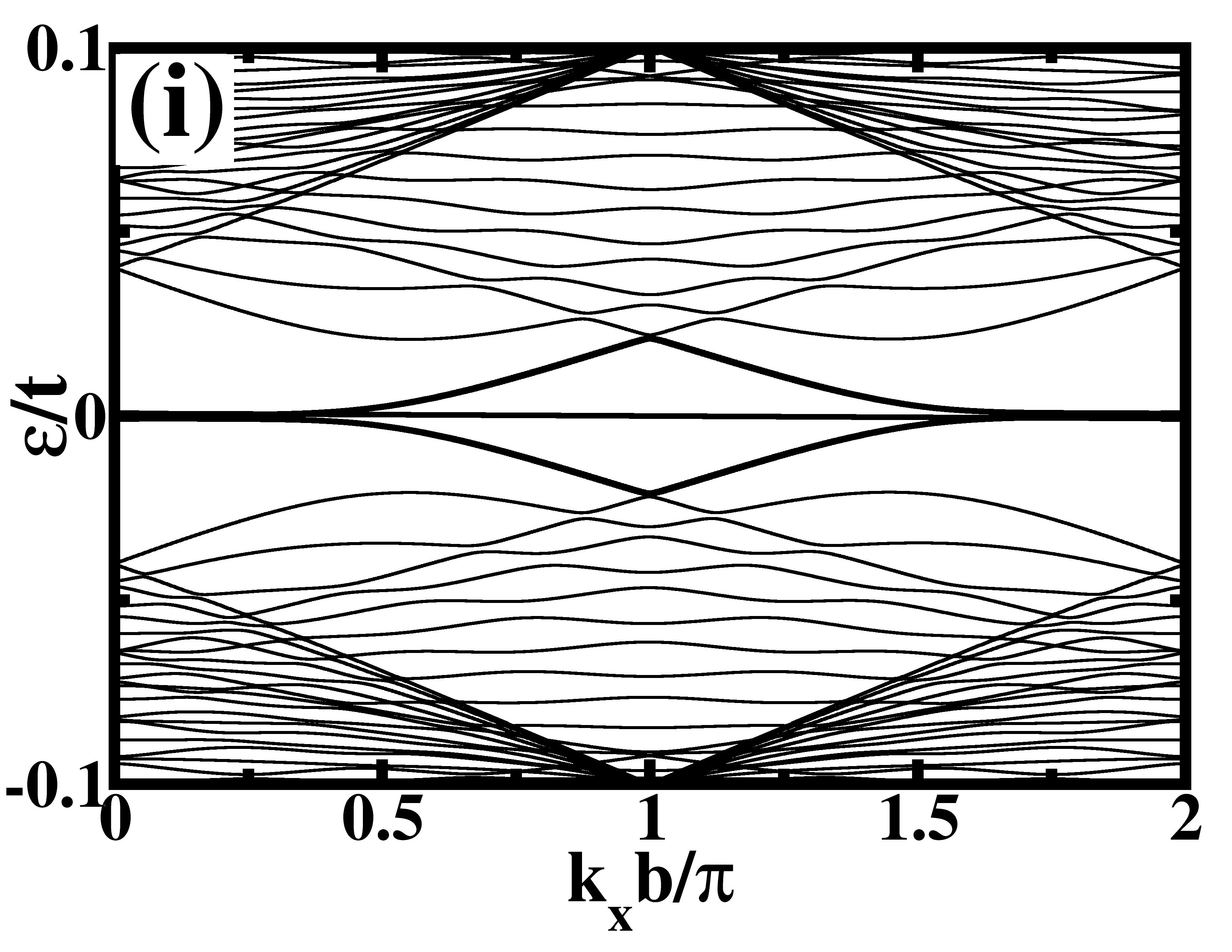}
	\caption{ Quasienergy spectra of pristine ZGNR(left panels), A-ZGNR
		(central panels) and B-ZGNR (right panels) are depicted under the influence of different field polarizations.
		(a), (b), and (c) correspond to LPL with $A_x = 2.56$, $A_y = 0$, and $\phi = 0$.
		(d), (e), and (f) correspond to LPL with $A_x = 0$, $A_y = 2.21$, and $\phi = 0$.
		(g), (h), and (i) correspond to CPL with $A_x = A_y = 2.21$ and $\phi = \pi/2$. The figures illustrate how the quasienergy spectrum of pristine ZGNRs, A-ZGNR and B-ZGNR responds to different field polarization and amplitudes in off-resonant limit providing insights into the electronic behavior of these systems.
	}
	\label{high_freq_band}
\end{figure*}

Figure \ref{monovacancy_band_normal}(a) illustrates the TB band structure of pristine ZGNR, while Figs. \ref{monovacancy_band_normal}(c) and \ref{monovacancy_band_normal}(e) depict the band structures of A-ZGNR and B-ZGNR, respectively, facilitating a comparative analysis of their band structures. In the case of pristine ZGNR, two degenerate localized bands emerge around the Fermi energy, while upon examining the band structures of A-ZGNR and B-ZGNR, a distinct single flat band appears at the Fermi energy. To delve deeper into these band structures, the electron density distribution (EDD) of edge atoms was plotted for both pristine and defected ZGNR. In Fig \ref{monovacancy_band_normal}(b), the 2-fold degenerate edge states in pristine ZGNR show contributions from both edge atoms. Figures \ref{monovacancy_band_normal}(d) and \ref{monovacancy_band_normal}(f) depict the EDD of edge atoms, illustrating the contribution of edge states to the flat band at the Fermi energy for A-ZGNR and B-ZGNR, respectively. These plots indicate that the defect-free edge significantly influences the flat bands. The elimination of a band at the Fermi energy in both A-ZGNR and B-ZGNR is a result of the pronounced contribution from the removed atoms within that particular band. Atoms along the defected edge contribute to bands situated slightly above and below the flat band occurred in the Fermi energy regime. In Fig. \ref{monovacancy_band_normal}(c) and \ref{monovacancy_band_normal}(e), it is evident that distinguishing between the band structures of A-ZGNR and B-ZGNR is quite challenging. Furthermore, Fig. \ref{monovacancy_band_normal}(a) is barely distinguishable from Fig. \ref{monovacancy_band_normal}(c) and \ref{monovacancy_band_normal}(e), indicating that the distinction between the band structures of pristine ZGNRs and defected ZGNRs is also minimal.
\begin{figure}[ht]
	\centering
	\includegraphics[width=4cm,height=4cm]{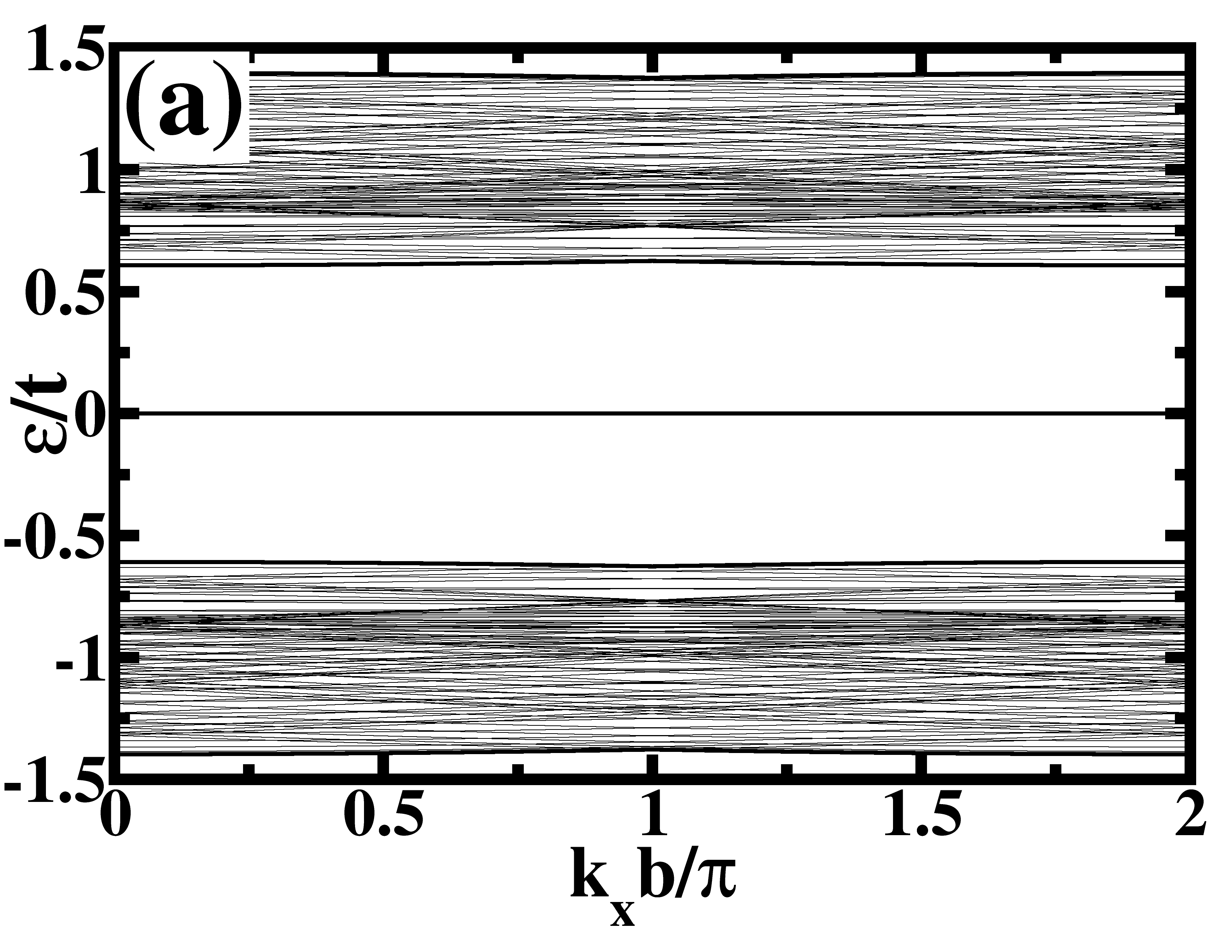}
	\includegraphics[width=4cm,height=4cm]{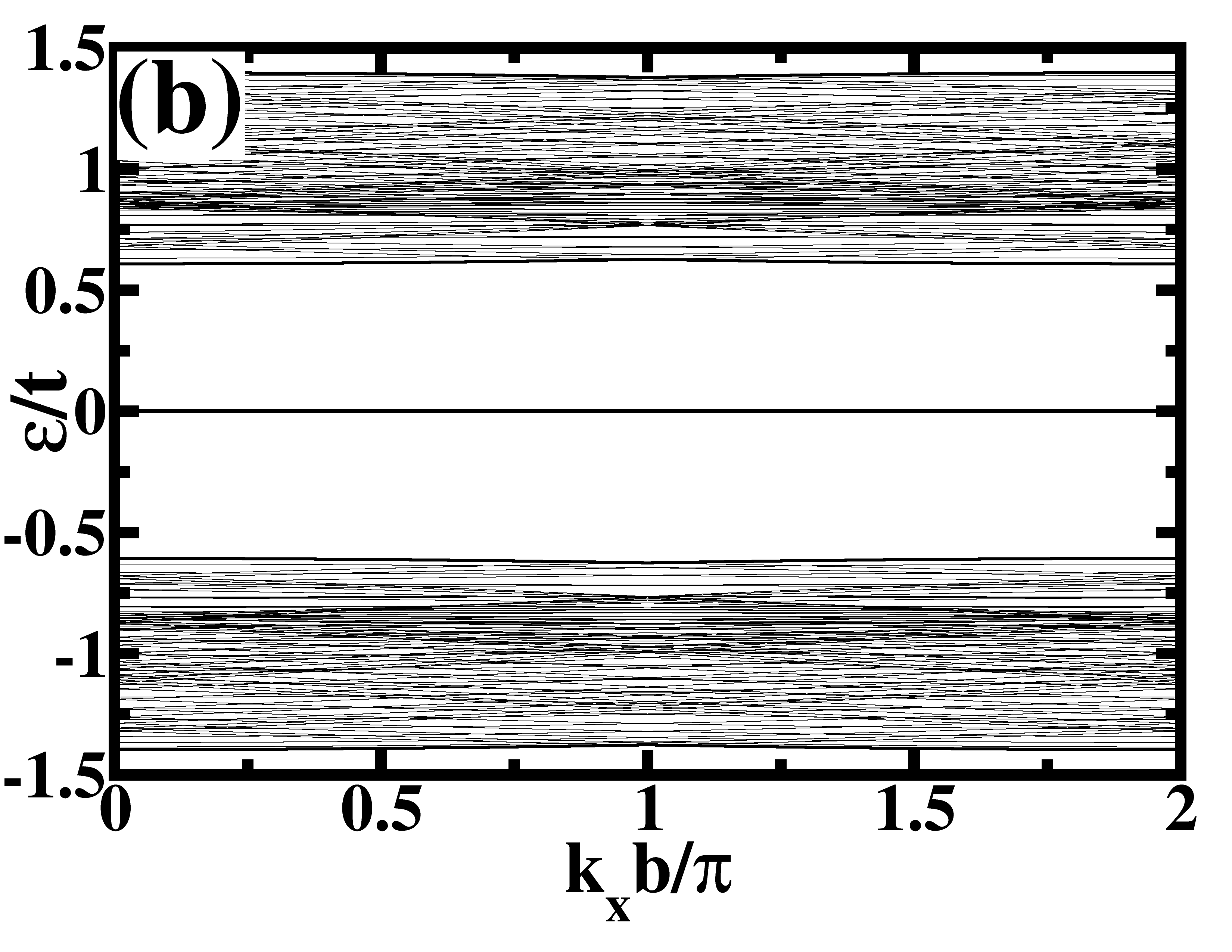}
	\includegraphics[width=4cm,height=4cm]{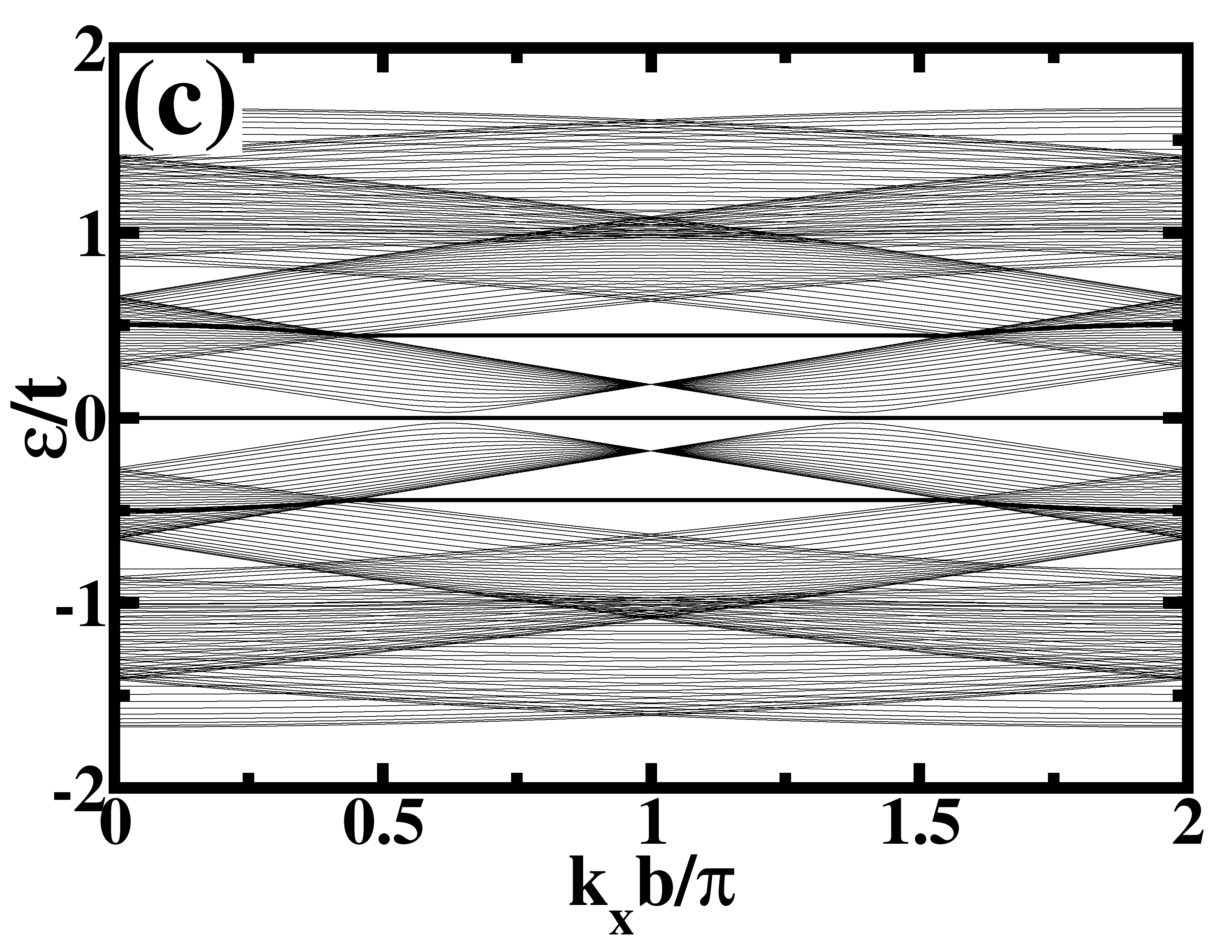}
	\includegraphics[width=4cm,height=4cm]{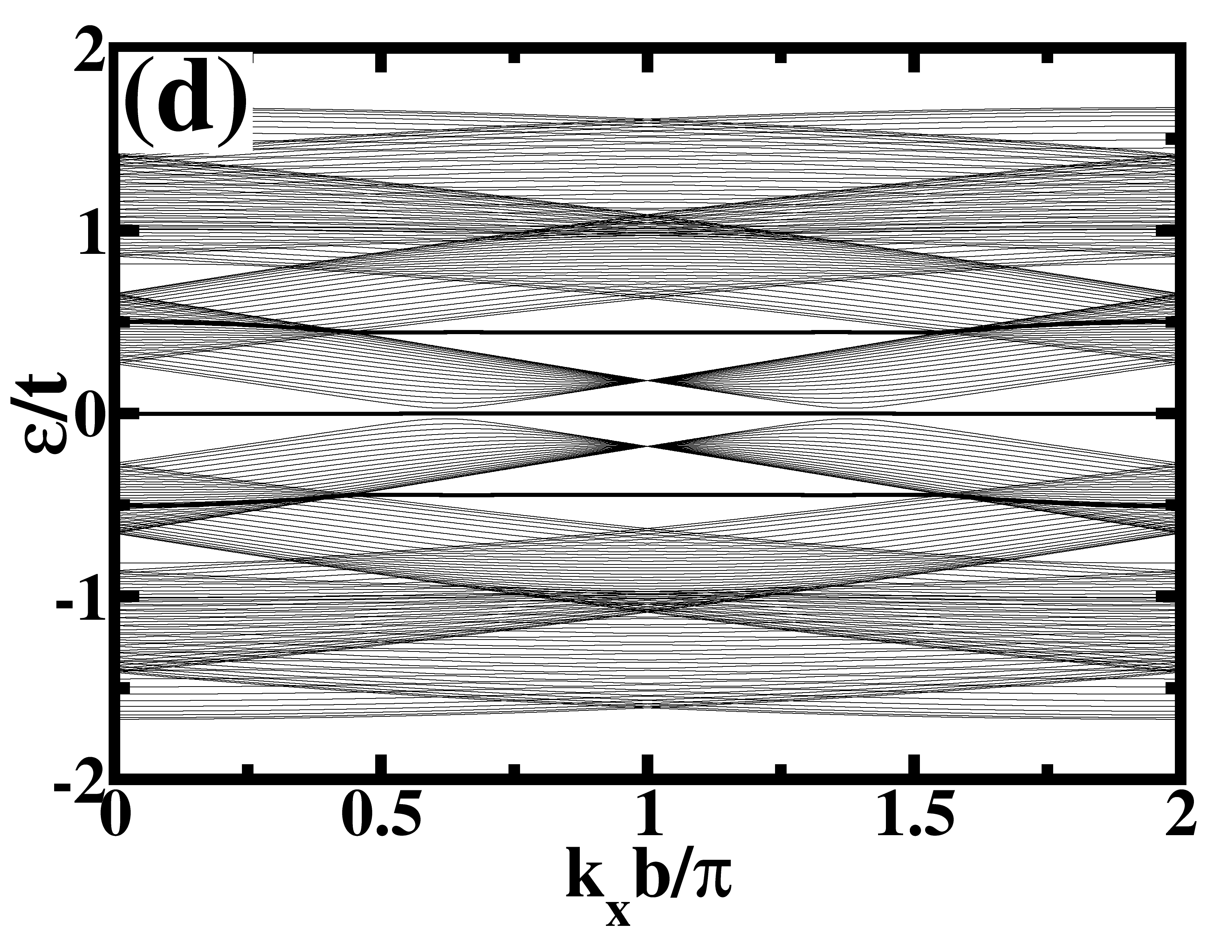}
	\caption{ Quasienergy spectra of A-ZGNR (left panels) and B-ZGNR (right panels) under the influence of LPL at $\omega = 6t$. (a) and (b) is for $A_x = 2.35$ and $A_y = 0$ while (c) and (d) is for $A_x = 0$ and $A_y = 2.04$ with phase $\phi = 0$. }
	\label{LPL_band}
\end{figure}
To address the challenges posed by the identical band structure of A-ZGNR and B-ZGNR, our aim is to manipulate the edge properties by employing light with diverse characteristics \cite{puviani2018time,calvo2012laser,wang2016edge,perez2014floquet}. By adjusting the parameters of the vector potential (${\bf A}(\tau)$), it becomes possible to introduce both linearly polarized light (LPL) and circularly polarized light (CPL). Modifying the amplitude ($A_{x,y}$), frequency ($\omega$), and phase ($\phi$) leads to alterations in the band structure. In off-resonant conditions, where the light frequency greatly exceeds the bandwidth ($\Delta W$), the Floquet bands remain independent, lacking inter-band coupling. The dominant term of the Floquet Hamiltonian is the zeroth Fourier component ($m = 0$). Consequently, Eq. (\ref{FEVE}) becomes block diagonal in Fourier space, consisting of a series of identical, time-independent Hamiltonians, each shifted by $\omega$. In these off-resonant driving scenarios, electronic states neither absorb nor emit photons due to the constraints of energy conservation. Rather, the electronic structures are modified through a process known as virtual photon exchange. This involves a combination of two second-order processes: one where an electron first emits a photon and then absorbs another photon, and another where an electron first absorbs a photon and then emits one \cite{virtual_phonon}. This mechanism leads to the dynamical localization of electrons, where electron hopping between sites can be effectively suppressed by carefully choosing the driving amplitude and frequency.
When the frequency is much lower than the bandwidth ($\omega \leq \Delta W$), a complex band structure emerges. The distinction between neighboring Floquet bands and their associated edge states becomes less apparent and more challenging to discern. However, as the frequency approaches the bandwidth ($\omega \sim \Delta W$), band crossings occur, leading to the formation of an energy gap and the potential emergence of edge states within it. In a physical sense, band crossing can be conceptually linked to processes involving one-photon or multiple-photon absorption and emission. This is why our focus is on the regime where the frequency is very high or comparable to the bandwidth.

Let us begin our discussion by exploring how light interacts with different polarization in the off-resonant (high-frequency) limit, focusing on pristine ZGNRs and ZGNRs with a monovacancy defect at the edges. In this regime, the interaction results in electron localization along specific orientations determined by the field polarization and intensity, introducing anisotropy into the system. This localization can be quantified through the renormalized hopping strength, denoted as $t_{j,0}^F$, where $j = 1,2,3$ represents nearest neighbors. The value of this hopping integral is influenced by the polarization and intensity of light. As the light intensity increases, the value of $t_{j,0}^F$ varies according to the Bessel function, indicating a damping effect. At certain intensity levels, the hopping along specific bonds is entirely suppressed, leading to electron localization \cite{Liu}. When linearly polarized light (LPL) travels along the x-direction (x-LPL) and interacts with the ZGNR, the hopping strength becomes $t_{1,0}^F = t$, while $t_{2,0}^F$ and $t_{3,0}^F$ vary with changing amplitude. Figure \ref{high_freq_band}(a) illustrates the quasienergy spectrum of pristine ZGNR for $A_x = 2.56$, which reduces the hopping strength by $t/10$ along $\delta_2$ and $\delta_3$. It is evident from the figure that due to electron localization along $\delta_1$, the quasienergy spectrum of the ZGNR resembles bulk atoms along $\delta_1$ behaving as carbon dimers, with corresponding bands appearing at $\epsilon = \pm t$, while edge atoms behave like isolated carbon atoms, resulting in bands appearing at $\epsilon = 0$. When A-ZGNR and B-ZGNR interact with light, the single removal of a carbon atom at the edge does not significantly impact the quasienergy spectrum, merely removing one flat band from $\epsilon = 0$, as depicted in Fig. \ref{high_freq_band}(b) for A-ZGNR and Fig. \ref{high_freq_band}(c) for B-ZGNR. For  $A_x = 2.584$, the hopping strength along $\delta_2$ and $\delta_3$ is completely suppressed, leading to the quasienergy of ZGNR portraying behavior akin to carbon dimers and isolated atoms (refer to Appendix C1 for further explanation). From Figs. \ref{high_freq_band}(a), \ref{high_freq_band}(b), and \ref{high_freq_band}(c), it is clear that detecting pristine ZGNRs and defected ZGNRs from quasienergy spectra using x-LPL in an off-resonant condition is not possible. Similarly, when a ZGNR is exposed to LPL along the y-direction (y-LPL), the hopping integrals along $\delta_1$ bonds change more rapidly compared to $\delta_2$ and $\delta_3$ bonds for certain amplitudes. Due to this variation in hopping strength, electrons start localizing equally along $\delta_2$ and $\delta_3$ bonds. In Fig. \ref{high_freq_band}(d), for instance, at $A_y = 2.21$, the quasienergy spectrum of ZGNR begins to exhibit characteristics resembling 1-D carbon chains. In the presence of a monovacancy defect at one edge of the ZGNR, the resulting effective 1D-carbon chain from the defected edge behaves as subchains, resembling small flakes of carbon chains. Figures \ref{high_freq_band}(e) and \ref{high_freq_band}(f) depict the band structure of A-ZGNR and B-ZGNR, respectively, under the influence of y-polarized light with an amplitude of $A_y = 2.21$, highlighting the unique features of 1D-carbon chains and carbon flakes. The quasienergy spectra of A-ZGNR and B-ZGNR suggest the presence of flat bands distributed along the energy axis, reminiscent of the band structure of carbon flakes consisting of edge carbon chains with monovacancies. For $A_x=2.23$, the hopping along the $\delta_1$ bond is entirely suppressed, causing the quasienergy spectrum of ZGNR to closely resemble that of a 1-D carbon chain, as depicted in Fig. \ref{fig:LPL_Ay_normal_very_high_omega}(d) of Appendix C. Again, with the assistance of y-LPL, we can distinguish only between pristine ZGNR and defected ZGNR, while discerning between A-ZGNR and B-ZGNR remains challenging in the off-resonant case. However, when a ZGNR is exposed to circularly polarized light (CPL) within the off-resonant limit, it undergoes a uniform change in hopping strength along the $\delta_1$, $\delta_2$, and $\delta_3$ bonds as the amplitude varies. As the amplitude increases, electron localization at individual carbon atoms intensifies. Consequently, all atoms within the ZGNR begin to behave like isolated entities, with their bands localizing at $\varepsilon = 0$ on the energy scale. Figures \ref{high_freq_band}(g), \ref{high_freq_band}(h), and \ref{high_freq_band}(i) represent the quasienergy spectrum of the pristine ZGNR, A-ZGNR, and B-ZGNR, respectively, for $A_x = A_y = 2.21$. From these figures, it is evident that bands start localizing at $\epsilon = 0$, indicating electron localization at respective atoms. The introduction of a monovacancy defect at either edge of the ZGNR has minimal impact, merely removing one band from the zero-energy level. When the amplitude of CPL is set to $A_x = A_y = 2.38$, the hopping along all nearest neighbors is completely suppressed, and the energy spectrum illustrates that the ZGNR behaves entirely like a collection of isolated atoms (for detailed explanation, see Appendix C3). Similar theoretical study for the localization of charge, where hopping between sites can be completely suppressed in the off-resonant limit, was conducted by Dunlap and Kenkre \cite{dunlap}, and Delplace et al. observe the localization of charge along 1D-carbon chain and at individual atoms of graphene \cite{merging}. From the preceding discussion on the off-resonant condition, we can conclude that discerning between pristine ZGNR, A-ZGNR, and B-ZGNR based on the quasienergy spectra for any polarization and amplitude of light is notably arduous.

Now, let's explore the regime where the frequency of light closely matches the bandwidth of a typical ZGNR, denoted as $\omega \sim \Delta W$. In this specific regime, the Floquet Hamiltonian doesn't effectively behave as a block diagonal matrix, and the interaction matrices (${H}_{m}$) become pivotal in shaping the quasienergy band structure. From the renormalized hopping integral, it becomes evident that the elements within these interaction blocks rely on higher-order Bessel functions with additional phase terms. These phase terms are contingent on the values of $A_x$, $A_y$, and $\phi$. We opt to set $\omega = 6t$ as our chosen frequency value, as at this frequency, the impact of the interaction term is significant. Moreover, this choice ensures that we can prevent any band crossings between the Floquet side bands for various amplitude values, simplifying our analysis. 
Figure \ref{LPL_band} presents the quasienergy spectrum of A-ZGNR (left panels) and B-ZGNR (right panels) under the influence of LPL. Panels \ref{LPL_band}(a) and \ref{LPL_band}(b) illustrate the quasienergy spectrum for $A_x = 2.35$ and $A_y = 0$, while panels \ref{LPL_band}(c) and \ref{LPL_band}(d) depict the scenario with $A_x = 0$ and $A_y = 2.04$ at $\phi = 0$. The quasienergy spectrum reveals similar trends observed in off-resonant limit, albeit with a subtle influence from the interaction terms. These interactions prevent any degeneracy in the bulk bands. It is discouraging to find that the quasienergy band structure remains entirely identical for both A-ZGNR and B-ZGNR when subjected to LPL along different polarization, regardless of the frequency regime. Consequently, whether considering ZGNRs with or without defects under the influence of LPL, the Floquet quasienergy spectrum exhibits an elegant inversion symmetry and notably manifests the presence of particle-hole symmetry (PHS).
\begin{figure}[ht]
	\centering
	\includegraphics[width=9cm,height=7cm]{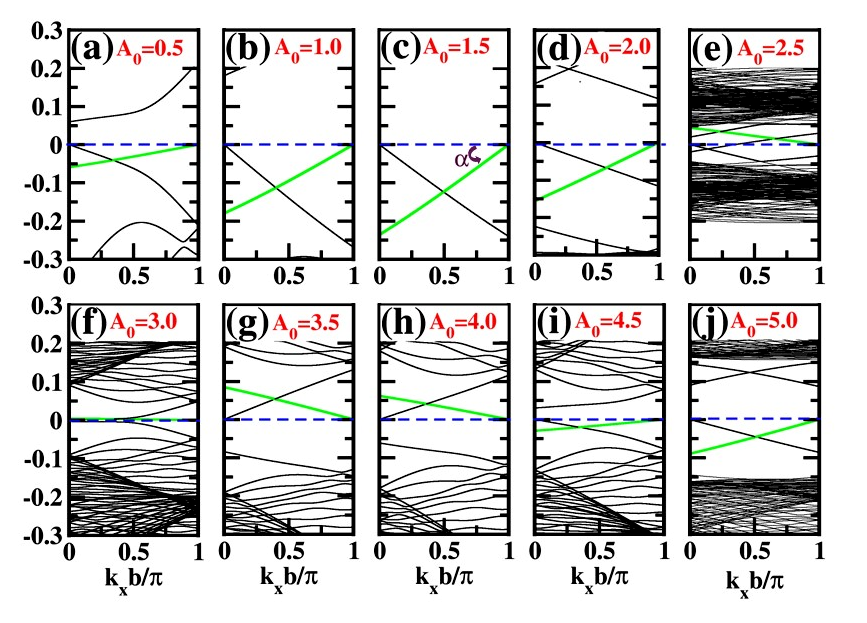}
	\caption{The quasienergy spectra of a A-ZGNRs with $N=20$ and $l = 5$, under the influence of left-circularly polarized light with a phase angle of $\phi = -\pi/2$ and frequency of $\omega = 6t$, while systematically varying the amplitude $A_0$ from 0.5 to 5 in increments of 0.5, as indicated by plots labeled from (a) to (j). The band colored in green corresponds to the edge without defect. $\alpha$ is the angle between the rotating band and $k_x$ axis. }
	\label{fig:cpl_varying_A}
\end{figure}

As we shift our attention to exploring the interaction of CPL with ZGNRs in the frequency regime comparable to bandwidth, it's important to note that extensive research has already delved into investigating the topological phases and chiral edge states, particularly concerning various field parameters. CPL introduces a unique aspect by inducing a rotation in its electric fields as they propagate through space. The distinct phase induced in interaction blocks due to this rotation varies for different nearest neighbors. This interaction triggers a phenomenon known as chiral symmetry breaking, selectively enhancing electron propagation along one edge direction while concurrently suppressing it along the opposite edge. This intricate process ultimately gives rise to the emergence of chiral edges.

At this stage, it is foreseeable that CPL can interact with the edge states localized within the zero-energy gap of the band structure in ZGNRs, especially when edge defects are present. Now we choose to focus on the quasienergy spectra of A-ZGNR as it interacts with CPL. Our initial approach involves systematically varying the amplitudes $A_x$ and $A_y$, particularly within a frequency regime comparable to bandwidth. We maintain $A_x = A_y = A_0$, where $A_0$ varies from $0.5$ to $5$ in increments of $0.5$. Additionally, we set the phase angle $\phi = -\pi/2$ to induce left-circularly polarized light. 
The interaction between the edge state and the applied periodic driving yields significant changes, as depicted in Fig. \ref{fig:cpl_varying_A}. Panels \ref{fig:cpl_varying_A}(a) to \ref{fig:cpl_varying_A}(c) illustrate an anticlockwise (ACW) increase in the angle ($\alpha$) between the edge state (highlighted in green) and the $k_x$ axis (indicated by the dashed violet line). Subsequently, as the amplitude ($A_0$) increases, this $\alpha$ decreases and eventually reaches zero, as evident in panels \ref{fig:cpl_varying_A}(d) through \ref{fig:cpl_varying_A}(f). Further increases in $A_0$ shift the direction of rotation to clockwise (CW), as demonstrated in panels \ref{fig:cpl_varying_A}(g) and \ref{fig:cpl_varying_A}(h). As we continue to increment $A_0$, it induces changes in the oscillation direction, leading to a motion resembling oscillating angular patterns around the point where $\varepsilon = 0$ and $k_x = \pi/b$.
A similar oscillatory behavior is observed in the band width, characterized by damped oscillations as the amplitude increases. This damping and oscillatory pattern arise due to the renormalization of the nearest-neighbor hopping parameter, given by $tJ_m(A_0a)$, with $J_m(A_0a)$ representing the $m^{th}$-order Bessel function. The inherent nature of Bessel functions to exhibit damping oscillations accounts for this phenomenon. A corresponding observation was made in a study involving a Kagome lattice ribbon by C. He et al. \cite{Kagome}
\begin{figure}[ht]
	\centering
	\includegraphics[width=4.25cm,height=5cm]{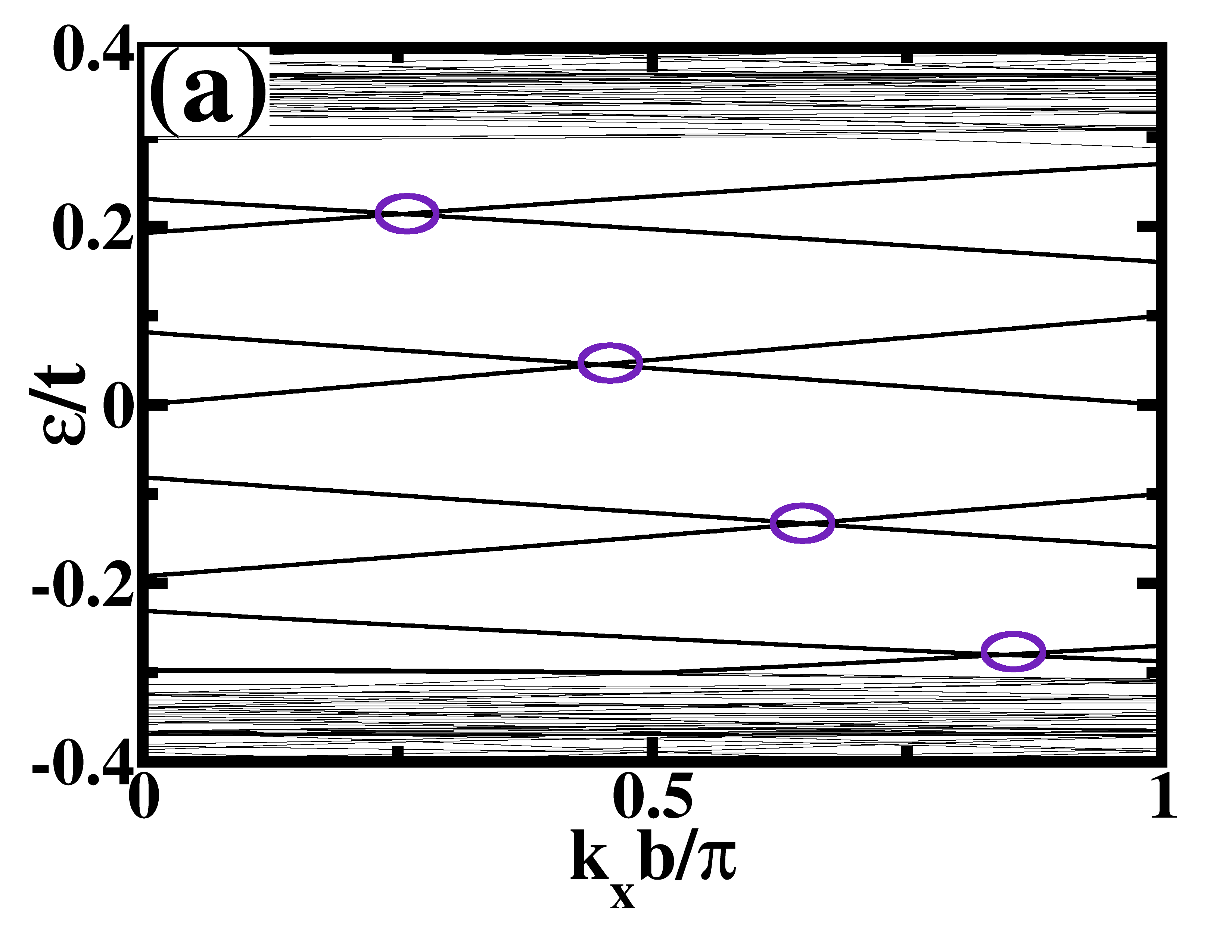}
	\includegraphics[width=4.25cm,height=5cm]{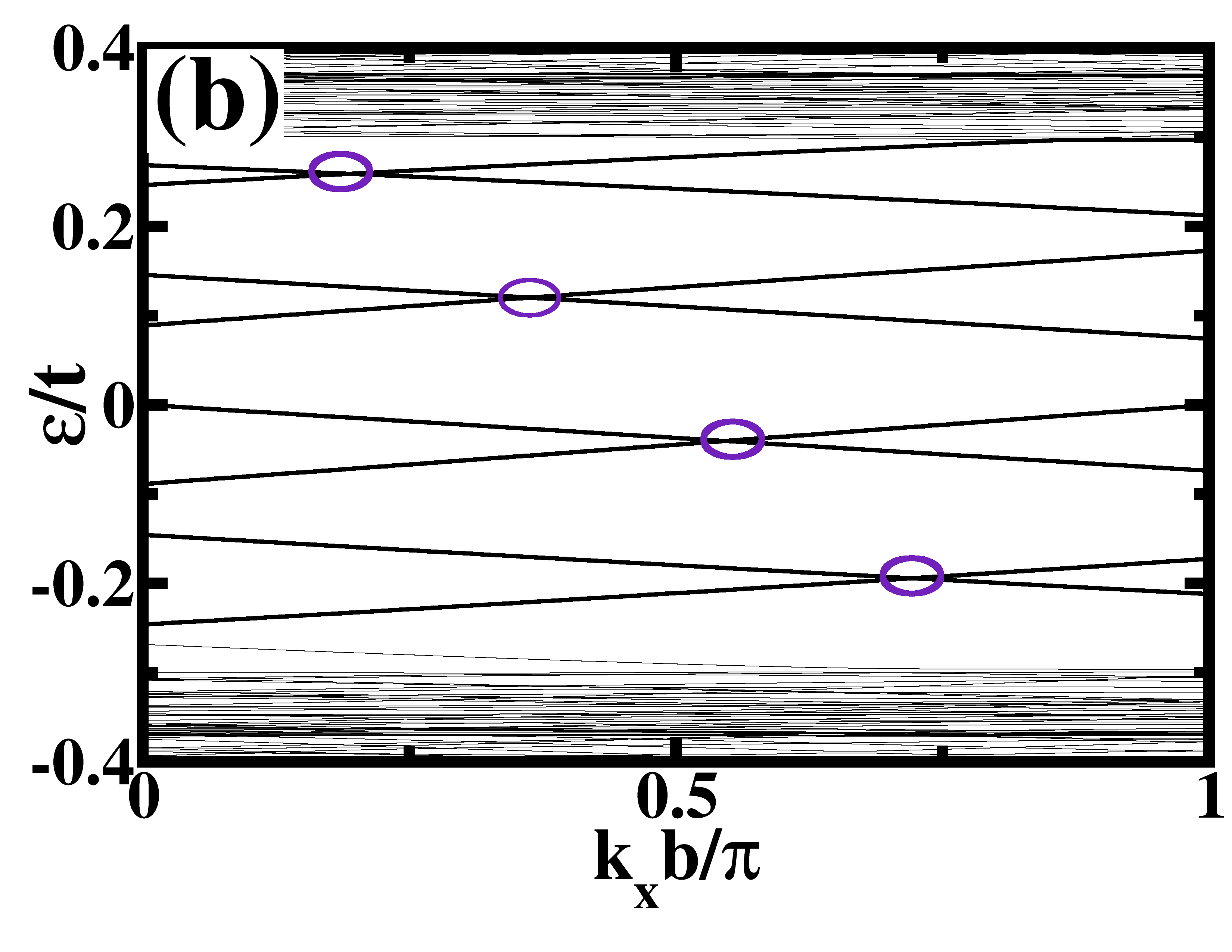}
	\caption{Qusienergy spectra of A-ZGNR for $A_0 = 2.04$ and $\phi = -\pi/2$ for (a)$l = 8$ and (b)$l = 9$. Circle indicates the Dirac points. }
	\label{fig:varying_l_8_9}
\end{figure}
At this stage, we have the flexibility to fine-tune the amplitude according to our preferences. We have chosen to set $A_0 = 2.04$, as this particular value provides an optimal angle for our observations. Additionally, we have opted for a lower amplitude to avoid the risk of significantly increased intensity, which could potentially harm the system.
Once we have established the amplitude, we also have the option to modify another optical parameter: the phase angle ($\phi$). It's important to note that adjusting the phase angle between $+\pi/2$ and $-\pi/2$ induces a shift in the polarization of light, transforming it from right-circularly polarized to left-circularly polarized. This alteration in polarization, coupled with the presence of the edge state, subsequently results in a change in the direction of oscillation.
\begin{table}[ht]
	\caption{\label{tab:summary_table}%
		Summary of observations for field phase variation, even/odd 'l' values, and ZGNR with monovacency defect at edge represented by A-ZGNR and B-ZGNR.
	}
	\begin{ruledtabular}
		\centering
		\begin{tabular}{c c c c}
			\textrm{Phase($\phi$)}&\textrm{Separation($l$)}&\textrm{Rotation}&
			\textrm{Monovacancy}\\
			\colrule
			$+\pi/2$ & 2n & ACW & A-ZGNR\\
			$+\pi/2$ & 2n & CW & B-ZGNR\\
			$+\pi/2$ & 2n+1 & ACW & B-ZGNR\\
			$+\pi/2$ & 2n+1 & CW & A-ZGNR\\
			$-\pi/2$ & 2n & ACW & B-ZGNR\\
			$-\pi/2$ & 2n & CW & A-ZGNR\\
			$-\pi/2$ & 2n+1 & ACW & A-ZGNR\\
			$-\pi/2$ & 2n+1 & CW & B-ZGNR\\
		\end{tabular}
	\end{ruledtabular}
\end{table}

Furthermore, our research has led us to examine how the width and size of the supercell of ZGNRs influence the rotation angle. At first, our investigation focused on observing the impact of the quasienergy spectra of A-ZGNR while varying the width (N) and keeping the supercell size constant. Interestingly, as we incrementally increased the width of the ribbon, we observed that the rotation angle remained unchanged. This observation is intriguing because altering the width of the A-ZGNR doesn't impact the edge state, as the number of atoms at the edge remains the same, while the bulk region becomes denser. This finding suggests that, even with the expansion of the width of the ribbon, the rotation angle remains consistent under fixed phase and amplitude conditions for CPL.
However, when we choose to adjust the spacing between the array of monovacencies along the edge, it necessitates a corresponding modification in the supercell size. The distance between these individual vacancies is determined by a parameter $l$, which determines the lattice vector of the supercell. As we increase the value of $l$, the separation between the arrays of defects at the edge widens. This widening gap gives rise to an increasing number of Dirac points due to the edge states nestled between the bulk valence and conduction bands. The reason behind this phenomenon lies in the fact that the band structure is dictated by the crystal lattice periodicity, which is described by the Brillouin zone. Altering the size of the supercell essentially entails changing the dimensions of the Brillouin zone. This can lead to shifts and distortions in the band structure as the allowed energy levels adapt to the new periodicity. We calibrate the value of $l$ with respect to the number of Dirac points (n) while keeping a fixed value of $A_0 = 2.04$. We have observed that for even values of $l$, the number of Dirac points appears to be $n = l/2$, whereas for odd values of $l$, it becomes $n = (l-1)/2$ (for explanation, see the Appendix C). We can determine the even-odd condition for $l$ by examining the maxima of the valence bulk band and the minima of the conduction bulk band associated with the bulk carbon atoms. For even values of $l$, these extrema occur at $k_x = 0$ whereas for odd values of $l$ they shift to $k_x = \pi$. Figure \ref{fig:varying_l_8_9} illustrates the quasienergy spectra of A-ZGNR, subject to modification by CPL with parameters $A_0 = 2.04$ and $\phi = -\pi/2$. Specifically, Fig. \ref{fig:varying_l_8_9}(a) corresponds to the case with an even value of $l = 8$, while Fig. \ref{fig:varying_l_8_9}(b) pertains to an odd value of $l = 9$. In both plots, the number of Dirac points remains the same, with $n = 4$.

The summarized observations, detailed in Table \ref{tab:summary_table}, provide insights into the rotational behaviour of localized edge states within the zero-energy gap region of the quasienergy spectra of A-ZGNR and B-ZGNR. These observations are categorized based on various factors, including the field phase ($\phi$), the `$l$' values (even or odd), and the position of the monovacancy defect at either the A-type or B-type edge of the ZGNRs. By examining these observations in the Table, we can make predictions regarding the location of the monovacancy defect at either of the edges. To achieve this, we utilize the known parameters of the incident light, namely the amplitude ($A_0$) and phase ($\phi$). When analyzing the quasienergy spectra, our initial step involves identifying the number of Dirac points, which, in turn, informs us about the value of $l$. Subsequently, we assess the rotation of the edge state around the point ($k_x = \pi, \varepsilon = 0$). These collective observations empower us to make accurate predictions regarding the precise location of the monovacancy defect within the edges of the structure.
\begin{figure}[ht]
	\centering
	\includegraphics[width=4.25cm,height=5.0cm]{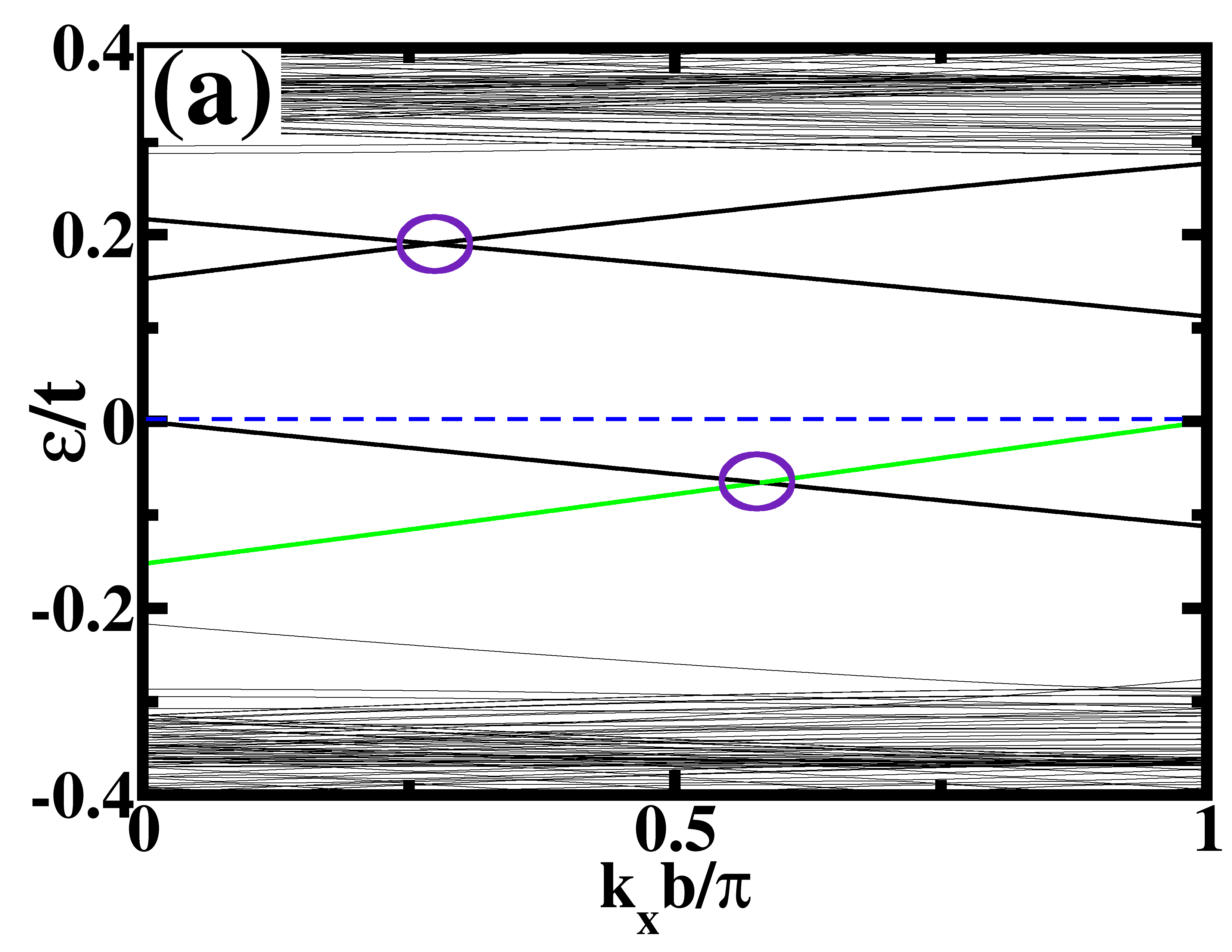}
	\includegraphics[width=4.25cm,height=5.0cm]{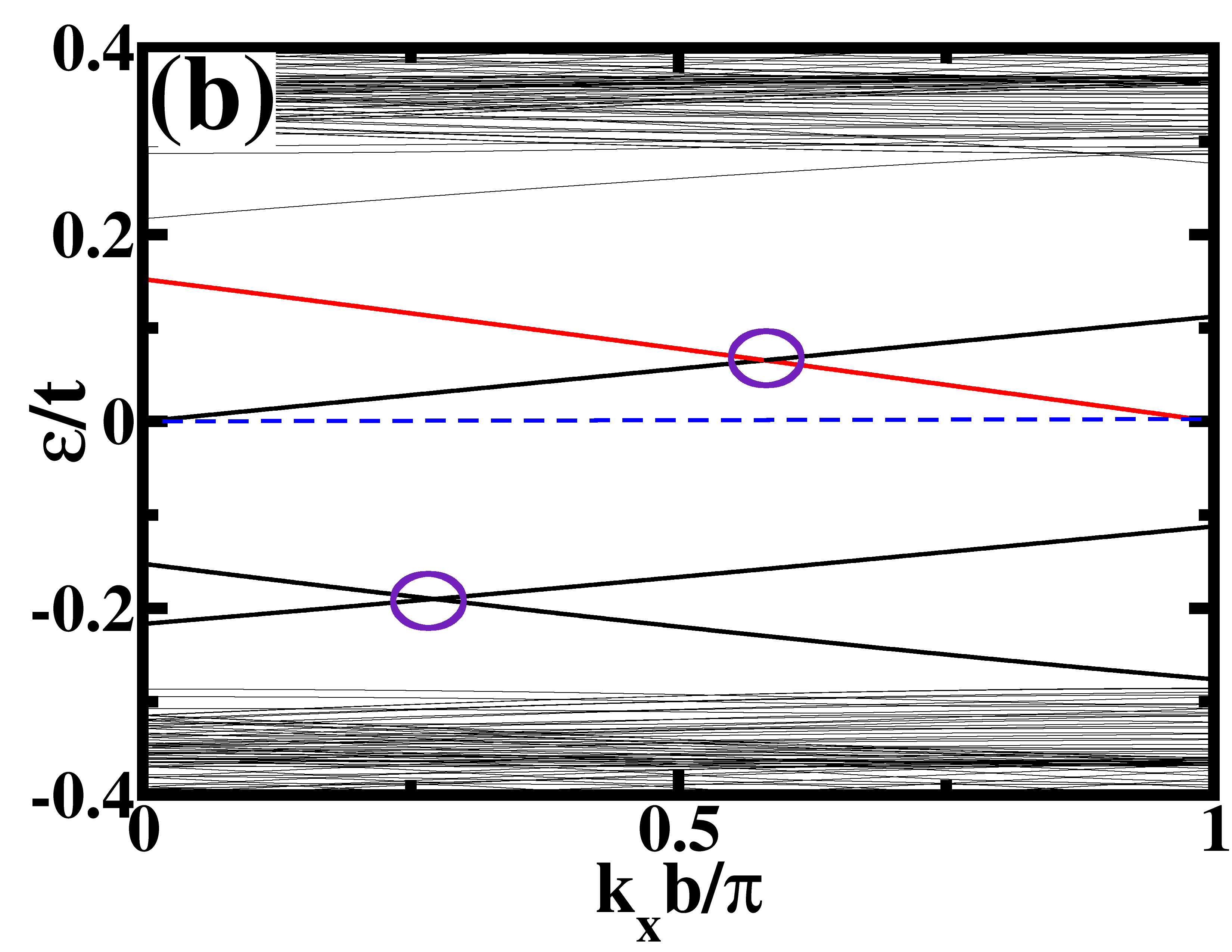}
	\caption{Quaienergy spectra of (a)A-ZGNR and (b)B-ZGNR  under the influence of CPL for $A_0 = 2.04$ and $\phi = -\pi/2$. All the Dirac points are encircled. Red color represents the edge of A-type and green color indicates the edge of B-type.}
	\label{fig:top_bottom}
\end{figure}

In this section, we conduct a comparative analysis of the quasienergy spectra of A-ZGNR and B-ZGNR, aligning with the earlier findings detailed in the table \ref{tab:summary_table}. Specifically, we set the amplitude of the illuminating light at $2.04$, with a phase angle of $\phi = -\pi/2$, signifying left-CPL. We carefully examine the quasienergy spectra presented in Fig. \ref{fig:top_bottom}.
Upon observation, we note that the rotation of the edge state depicted in Fig. \ref{fig:top_bottom}(a) is anticlockwise (ACW), while in Fig. \ref{fig:top_bottom}(b), it is clockwise (CW). Additionally, we observe that the number of Dirac points in both cases is $n = 2$. Given that the extremities of the bulk valence and conduction bands occur at $k_x = \pi$, it indicates that the value of $l$ should be odd.
By employing the formula for odd values of $l$, we determine that the value of $l$ is 5. Comparing this observed data with the information provided in the table, we can conclude that Fig. \ref{fig:top_bottom}(a) corresponds to a A-ZGNR, and Fig. \ref{fig:top_bottom}(b) corresponds to a B-ZGNR. The separation between the array of monovacancy defects at the edge is $l = 5$.

\section{Conclusion and Outlook}
In this comprehensive study, we have meticulously explored the intricate characteristics of the band structures of A-ZGNR and B-ZGNR. Our comparative analysis has revealed that the introduction of these monovacancy defects at the edges induces subtle yet significant changes in the band structures of ZGNRs.
These changes manifest in the emergence of flat bands within the zero-energy gap, which are notably influenced by the adjacent defect-free edges. We have also delved into the intriguing interactions between ZGNRs and both LPL and CPL across different frequency regimes. In the off-resonant limit, we have noted that light tends to localize charge along carbon-carbon bonds depending on the polarization of light and intensity. This process of localization can mimic the behavior of isolated carbon atoms, carbon dimers, single-atom carbon chains, or carbon flakes. When the frequency regime aligns with the band width, the interaction terms (${\bf H}_{m=\pm 1,2,3..}$) play a significant role in shaping the quasienergy spectra. In the case of LPL, our investigation revealed trends similar to those observed in off-resonant limit, albeit with subtle influences from the interaction terms. These interactions effectively prevent degeneracy in the bulk bands. A noteworthy discovery is that the quasienergy band structure remains entirely consistent for both A-ZGNR and B-ZGNR when subjected to LPL along both directions, regardless of the frequency regime.
Remarkably, in this regime, CPL interactions with ZGNRs lead to the breaking of chiral symmetry, resulting in the emergence of chiral edge states. Our research meticulously dissects the complex interplay between light orientation, magnitude, and the resulting behaviors of edge states, underscoring the sensitivity of edge state rotation angles to variations in light parameters.
Furthermore, our study has revealed that altering the width of ZGNRs has no impact on the rotation angles of edge states. In contrast, adjusting defect spacing plays a pivotal role in the formation of Dirac points—a critical finding concisely summarized in a comprehensive Table \ref{tab:summary_table}. This table elucidates the influence of distinct field phases, defect placements, and lattice dimensions on the rotation of edge states.
In brief, our research culminates in a thorough investigation that compares the quasienergy spectra of A-ZGNR and B-ZGNR. The findings consistently demonstrate that the associated flat band exhibit opposite rotation directions when subjected to CPL, aligning with our earlier observations. These insights allow us to predict the location of monovacency at either of the edges within ZGNRs.

While the experimental observation of the distinctive features of light-induced Floquet states in materials poses significant challenges, time- and angle-resolved photoemission spectroscopy (tr-ARPES) emerges as a particularly effective tool for delving into the intricacies of Floquet physics, especially in driven ZGNRs. tr-ARPES allows for the direct investigation of the dressed electronic states induced by external driving, providing a promising avenue for empirical exploration. S. Zhou et al. employed tr-ARPES in experimental investigations on black phosphorous, selected as a model semiconductor \cite{trARPES1}. Their study revealed experimental manifestations of Floquet band engineering. Additionally, Y. H. Wang et al. demonstrated that the tr-ARPES spectrum resulting from an intense ultrashort mid-infrared pulse, with energy below the bulk band gap, undergoes hybridization with the surface Dirac fermions of a topological insulator, leading to the formation of Floquet-Bloch bands \cite{trARPES2}. Furthermore, in a separate study, S. Aeschlimann et al. combined tr-ARPES with the time-dependent density functional theory of $WSe_2$ and graphene, aiming to explore the persistence of Floquet-Bloch states in the presence of scattering \cite{trARPES3}. Very recently, we come to know about direct observation of Floquet-Bloch states in monolayer graphene through tr-ARPES with mid-infrared pump excitation \cite{trARPES4}. Hence, by harnessing this theoretical knowledge, there is a possibility to experimentally demonstrate the quasienergy spectrum of A-ZGNRs and B-ZGNRs. This approach could be instrumental in detecting specific structural features, such as monovacancy defects, thereby bridging the gap between theoretical understanding and empirical realization in the realm of Floquet physics in graphene-based materials.

In essence, our research offers a deep comprehension of the complex dynamics between sub-lattice defects, external light interactions, and the consequent behaviors of edge states within ZGNRs. Consequently, this investigation offers valuable insights into comprehending the Floquet quasienergy spectra of both pristine ZGNRs and those with edge defects, opens up a door to utilizing Floquet theory as a tool for detecting defects in various systems by analyzing quasienergy spectra.
\section*{ACKNOWLEDGMENTS}
P. Parida acknowledges DST-SERB for the ECRA project (No. ECR/2017/003305). A. Kumar thanks University Grants Commission (UGC), New Delhi, Government of India, for financial support in the form of a Senior Research Fellowship (DEC18-512569-ACTIVE).
\appendix
\section{Floquet theory: ZGNR in ac field}
To compute the Floquet operator, we begin by examining the tight-binding Hamiltonian of the undriven ZGNR in reciprocal space ${\bf H}(k) = t\sum_{k}C_{k}^{\dag}C_{k}e^{i{\bf k.b}}$, employing the nearest-neighbor approximation, where $C_{\bf k}^{\dag}$ and $C_{\bf k}$ are the creation and annihilation operator, respectively. The width of the ZGNR is determined by selecting the number of carbon chains, denoted as N. The basis of the Hamiltonian relies on the lattice within the supercell. Consequently, the basis of the Hamiltonian is expressed as $2Nl$, where $l$ represents the number of unitcells in the supercell.
The influence of the ac field is introduced through the vector potential ${\bf A}(\tau)$, incorporating a time-dependent phase factor in the hoppings via Peierls substitution: $t \rightarrow t_j(\tau) = t e^{i{\bf A}(\tau) \cdot {\bf \delta}_j}$. Here, ${\bf \delta}_1 = a(0,1)$, ${\bf \delta}_2 = a/2(-\sqrt{3},-1)$, and ${\bf \delta}_3 = a/2(\sqrt{3},-1)$ represent the nearest neighbor vectors. 
 
The composed scalar product of the Floquet operator, denoted as ${\bf H}_F({\bf k},\tau) = {\bf H}({\bf k},\tau) - i\hbar\pdv{}{\tau}$, is calculated in the same space of time-independent basis $\ket{u_{\alpha,{\bf k},p}}$ as depicted in the eigenvalue equation in the main text. Solving Fourier component leads to the expression 
\begin{equation}
\begin{split}
    \frac{1}{2\pi}\int_{0}^{2\pi}e^{in\theta} e^{\eta(\phi)\cos(\theta)}e^{i\gamma(\phi)\sin(\theta)} d\theta = \\
    \frac{1}{2\pi}\int_{0}^{2\pi}e^{in\theta} e^{\eta(\phi)\cos(\theta)} \sum_{s = -\infty}^{\infty}J_{s}(\gamma(\phi))e^{is\theta}d\theta
    \end{split}
\label{jacobi_use}
\end{equation}
which can be solved using Jacobi-Anger expansion,
\begin{equation}
e^{\pm iz\sin(\theta)} = \sum_{s = -\infty}^{\infty}J_{s}(z)e^{\pm is\theta}
\label{jacobi_angar}
\end{equation}
where $J_{s}(z)$ is the Bessel function of $s^{th}$ order with argument $z$. On further solving leads to summation which can be solved by using identity
\begin{equation}
	\centering
\sum_{s = -\infty}^{\infty} J_{n+s}(\eta(\phi))J_{s}(\gamma{\phi})e^{is\phi} = e^{in\Lambda}J_n(\Delta)
\end{equation}

where ac field parameters are encoded in $\Lambda$ and $\Delta$ given by  

\begin{align}
\centering
    \Lambda &= \tan^{-1}\left(\frac{\gamma(\phi)\sin(\phi)}{\eta(\phi)-\gamma{\phi}\cos(\phi)}\right)\\
    \Delta &= \sqrt{\eta(\phi)^2+\gamma(\phi)^2-2\eta(\phi)\gamma(\phi)\cos(\phi)}
\end{align}
The detailed derivation of the Floquet operator in Sambe space is done by P. Delplace et al. \cite{ref37}

\section{Computational Details}
The formation energy of the monovacancy defect was determined through computational
	 density functional theory calculations to verify the stability of edge monovacancy in zigzag
	 nanoribbons. The geometry of the monovacancy zigzag nanoribbon was studied using the Vienna ab	 initio simulation package (VASP) with the projector augmented wave (PAW) method. The electron 
	 exchange and correlation potential have been described using the Perdew–Burke–Ernzerhof (PBE) 
	 functional within the generalized gradient approximation (GGA) framework. A cell size of $12.46 \times 30 \times 20.0 \AA$ with an energy cut-off of $600 eV$ was employed in the calculations. A vacuum 
	 separation distance of $30 \AA$ and $20 \AA$ along the y- and z-directions has been implemented to 
	 mitigate interactions beyond the periodic boundary conditions. Atomic positions are iteratively 
	 optimized until the energy change becomes less than $1 \times 10^{- 6} eV$ per cell and the force on each atom fell below $0.01 eV \AA^{- 1}$. K-point sampling for Brillouin zone
	 integration was carried out using a $6 \times 1 \times 1$ Monkhorst-Pack k-point mesh to
	 capture energy and electronic properties accurately.
 \begin{figure}[ht]
 	\centering
 	\includegraphics[width=8.5cm,height=3.5cm]{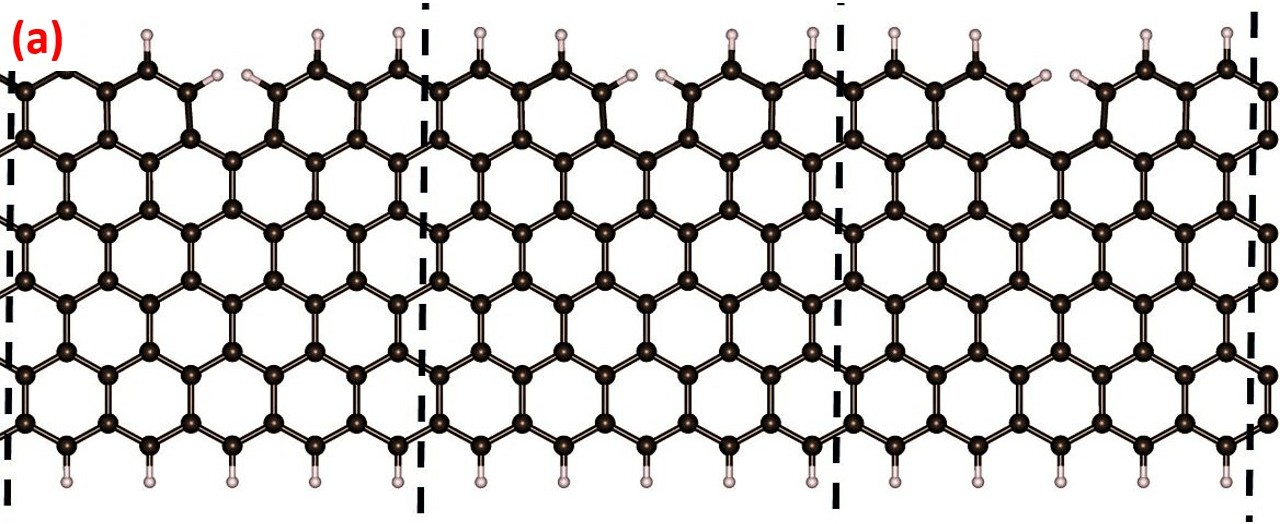}
 	\includegraphics[width=8.5cm,height=3.5cm]{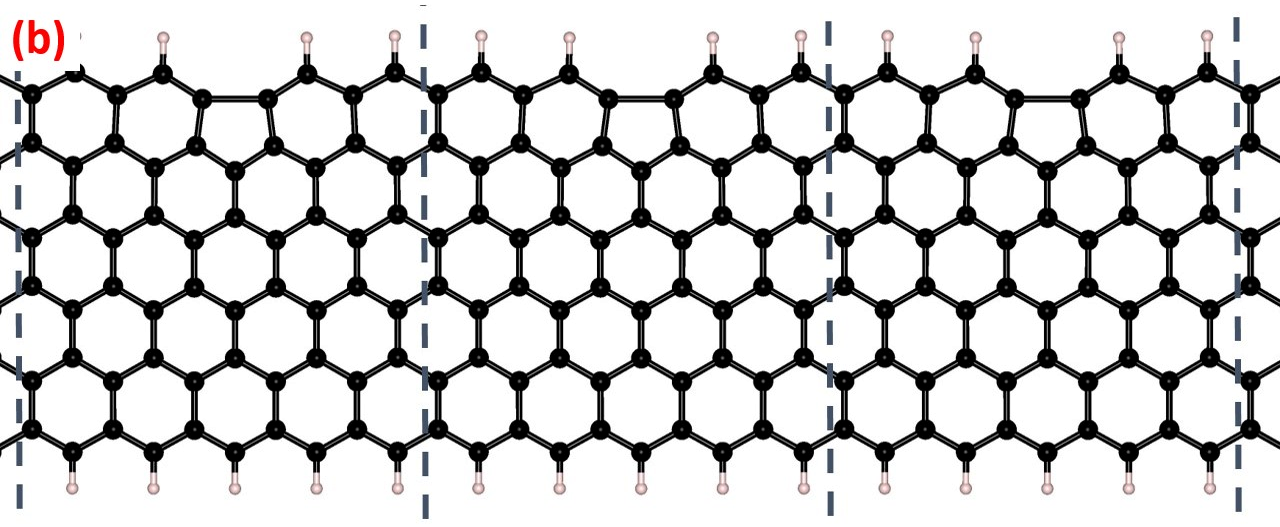}
 	\caption{Demonstration of ZGNR with (a) array of monovacancy defects at edge with optimized formation energy and (b) array of pentagon reconstruction at edge.}
 	\label{fig:dft_mv}
 \end{figure}
A unit cell of nanoribbons with a lattice constant of $|b| = 12.426 \AA$, periodic along the zigzag (x-axis) direction as shown in Fig. \ref{fig:dft_mv} . The edge monovacancy arises from the removal of a single carbon atom, with hydrogen atoms then stabilizing the resulting dangling bonds. Meanwhile, the formation of a pentagon occurs through the rearrangement of these dangling bonds caused by the creation of the monovacancy defect. Figure \ref{fig:dft_mv} illustrates the optimized geometries of both the monovacancy defect and the pentagon in the ZGNR at the edge.

\section{Interaction of ZGNR with light of different polarization in off-resonant condition.}
    In our study, we have selected zigzag graphene nanoribbons (ZGNRs) with a width composed of twenty carbon chains ($N = 20$), and the supercell consists of five unit cells ($l = 5$).
\subsection{Linear polarization along x-direction} When ZGNR is exposed to LPL radiation in the x direction, with ${\bf A} = (A_x \cos(\omega t),0)$, and the frequency is high ($\hbar\omega >> 6t$), the Floquet Hamiltonian has a structure that looks like a block-diagonal. Each block matrix is analogous to the undriven graphene Hamiltonian but experiences an eigenvalue shift of $\omega$. The renormalized hopping integral in the ${\bf\delta}_1$ direction remains constant regardless of the field amplitude. However, the hopping integrals along ${\bf\delta}_2$ and ${\bf\delta}_3$ are equivalent and exhibit variation with field amplitude, as expressed in the equation below.
\begin{equation}
\begin{split}
    t_1^F &= t J_0(0)\\
    t_2^F &= t J_0(\sqrt{3}A_xa/2)\\
    t_3^F &= t J_0(\sqrt{3}A_xa/2)
\end{split}\label{lpl_Ax_hopping}
\end{equation}
Examining the renormalized hopping term reveals that in the x-direction, the LPL exerts localization of charge on the ZGNR specifically affecting the ${\bf \delta}_2$ and $\delta_3$ bonds. As the field amplitude increases, the hopping strengths, denoted as $t_2^F$ and $t_3^F$, weaken, restricting electron mobility along the bonds. The preferred direction for electron hopping is predominantly along the $\delta_1$ bond.
\begin{figure}[ht]
	\centering
	\includegraphics[width=8.5cm,height=8.5cm]{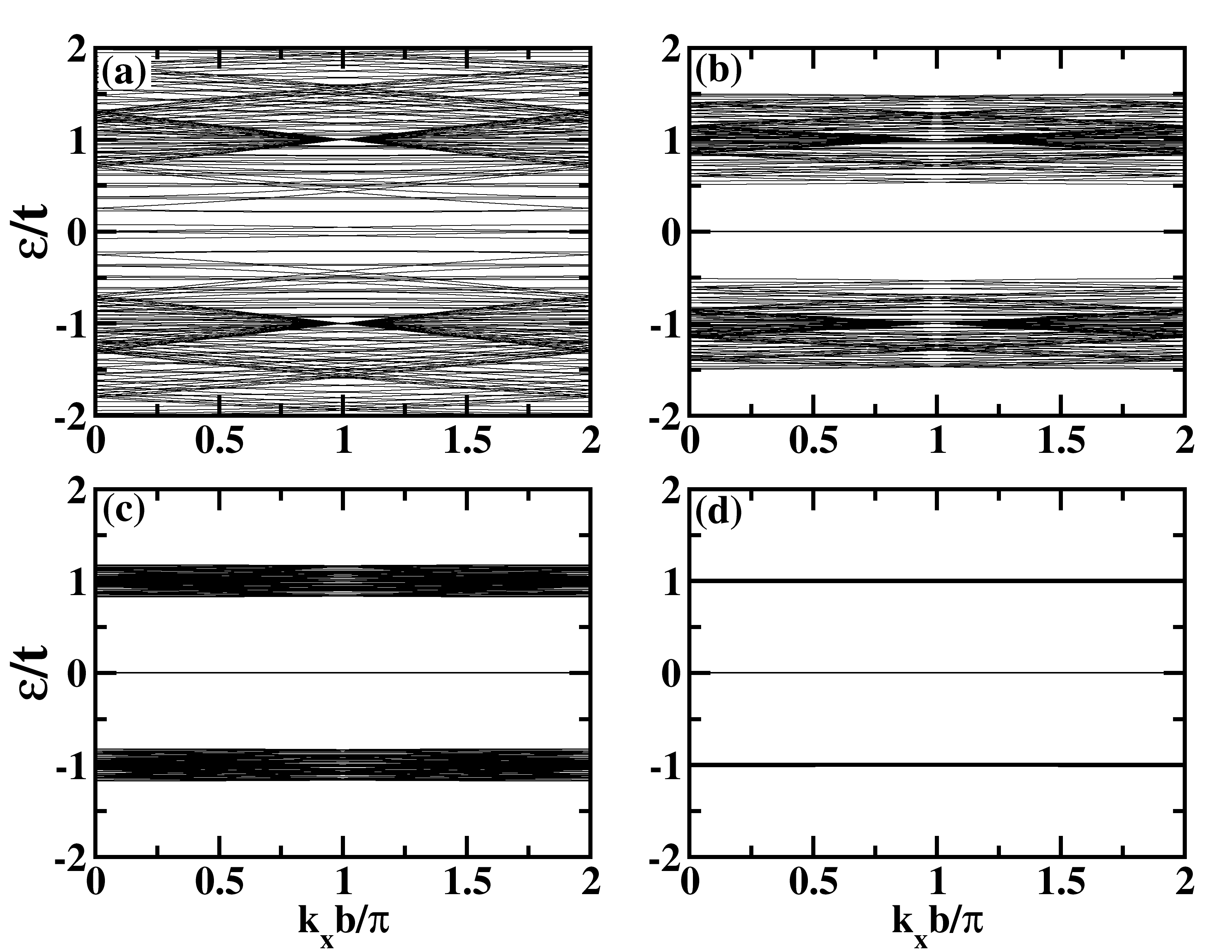}
	\caption{ZGNR irradiated with LPL along x-direction in high-frequency regime with (a)$A_x = 1.756$, (b)$A_x = 2.256$, (c)$A_x = 2.561$ and (d)$A_x = 2.584$ .}
	\label{fig:LPL_varying_Ax_very_high_omega}
\end{figure}
With a very high field strength, electron hopping along $\delta_2$ and $\delta_3$ bonds is entirely 
suppressed, and electrons primarily move along $\delta_1$. The quasi-energy band structure for this 
system is depicted in Fig. \ref{fig:LPL_varying_Ax_very_high_omega}. Figure
\ref{fig:LPL_varying_Ax_very_high_omega}(a) corresponds to $Ax = 1.756$, where the hopping strength 
$t_2^F$ and $t_3^F$ is reduced to half compared to $t_1^F$. Similarly, for Fig.
\ref{fig:LPL_varying_Ax_very_high_omega}(b), $A_x = 2.256$, resulting in a quarter of the hopping strength
along $\delta_2$ and $\delta_3$. Panels (c) and (d) of Fig. \ref{fig:LPL_varying_Ax_very_high_omega}
correspond to $A_x = 2.561$ and $A_x = 2.584$, respectively. In panels (c) and (d), the hopping
strengths $t_2^F$ and $t_3^F$ are reduced to one-tenth and one-hundredth of $t_1^F$, respectively.

Figure \ref{fig:LPL_varying_Ax_very_high_omega}(d) reveals the presence of ten flat bands at zero energy throughout the entire Brillouin Zone (BZ). This occurrence results from the electron localization at the edge atoms because edge atoms behaves as isolated atoms. Additionally, the quasienergy bands within the valence and conduction bands, stemming from the bulk atoms, become localized at $\varepsilon = \pm t$. This band localization phenomenon arises due to the behavior of bulk atoms resembling a carbon dimers configured along the $\delta_1$ direction. 

\subsection{Linear polarization along y-direction} In the high-frequency regime, when a ZGNR interacts with a LPL along the y-direction, the resulting effective renormalized hopping potential is described in Eq. \ref{lpl_Ay_hopping}. 
\begin{equation}
\begin{split}
    t_1^F &= t J_0(A_ya)\\
    t_2^F &= t J_0(A_ya/2)\\
    t_3^F &= t J_0(A_ya/2)
\end{split}\label{lpl_Ay_hopping}
\end{equation}

\begin{figure}[ht]
	\centering
	\includegraphics[width=8.5cm,height=8.5cm]{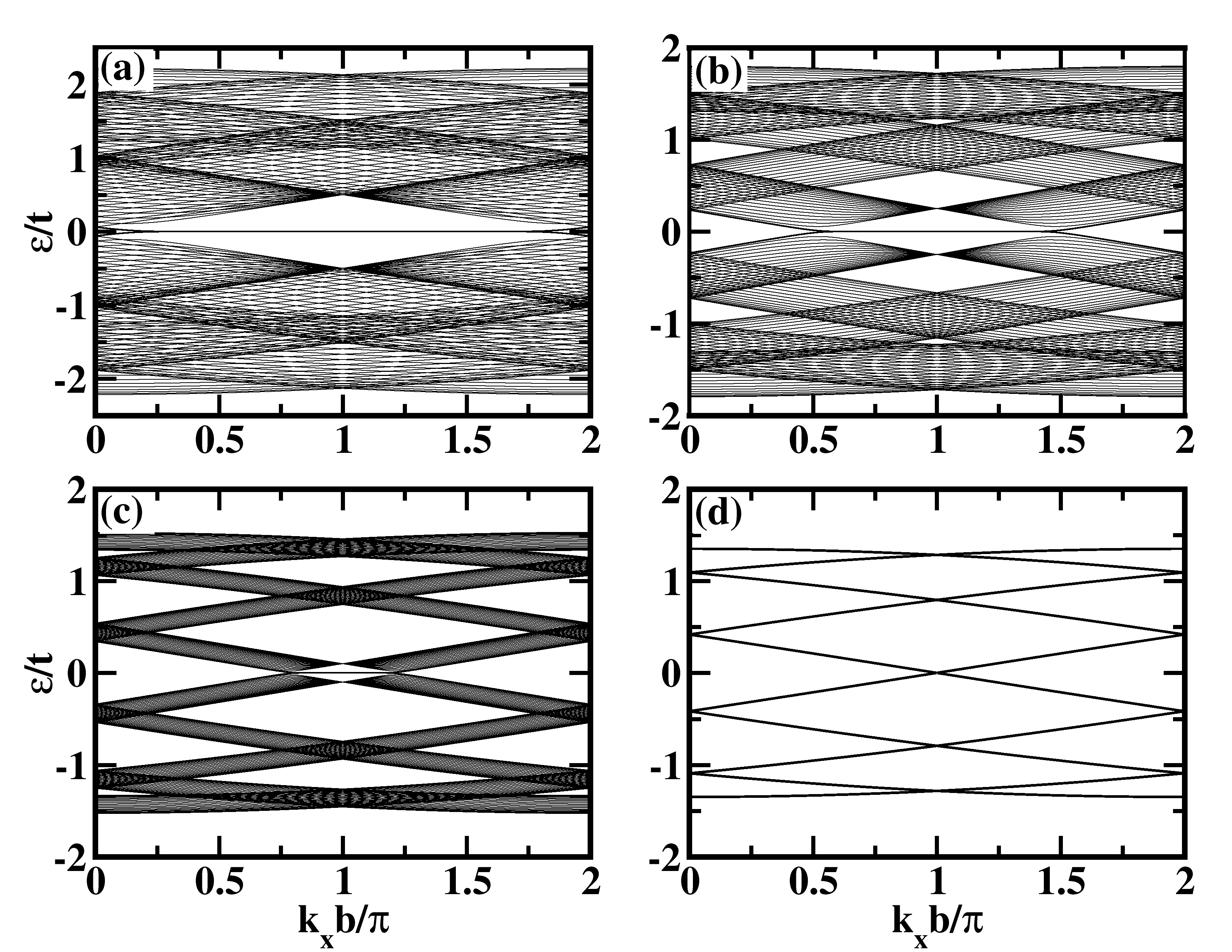}
	\caption{ZGNR irradiated with LPL along x-direction in high-frequency regime with (a)$A_y = 1.521$, (b)$A_y = 1.954$, (c)$A_y = 2.218$ and (d)$A_y = 2.238$ .}
	\label{fig:LPL_Ay_normal_very_high_omega}
\end{figure}

As the amplitude $A_y$ increases, it induces charge localization in the ZGNR in such a manner that the values of $\delta_1$, $\delta_2$, and $\delta_3$ decrease. However, the condition is maintained such that the value of the effective renormalized hopping energy, denoted as $t_1^F$, always remains greater than $t_2^F$ and $t_3^F$. This results in the localization of electrons along each carbon chain within the ZGNR, effectively causing each carbon chain in the ZGNR to exhibit properties akin to an infinite 1-D carbon chain.   

Figure \ref{fig:LPL_Ay_normal_very_high_omega} illustrates the quasienergy spectra of ZGNR for various values of $A_y$. Notably, Fig. \ref{fig:LPL_Ay_normal_very_high_omega}(d) depicts the band structure of ZGNR when $A_y = 2.238$, during which $t_1^F = t/100$, and $t_2^F = t_3^F$  is adjusted accordingly. At a specific value of $A_y$, the $t_1^F$ value reaches zero. In this scenario, each chain within the ZGNR becomes completely isolated from the others, behaving as an independent 1-D carbon chain.
\subsection{Circularly polarised light} When the parameters of light set to $A_x = A_y = A_0$ and $\phi = \pm \pi/2$, then hopping integral in high-frequency regime is given in Eq. \ref{cpl_hopping}.
\begin{figure}[ht]
    \centering
    \includegraphics[width=8.5cm,height=8.5cm]{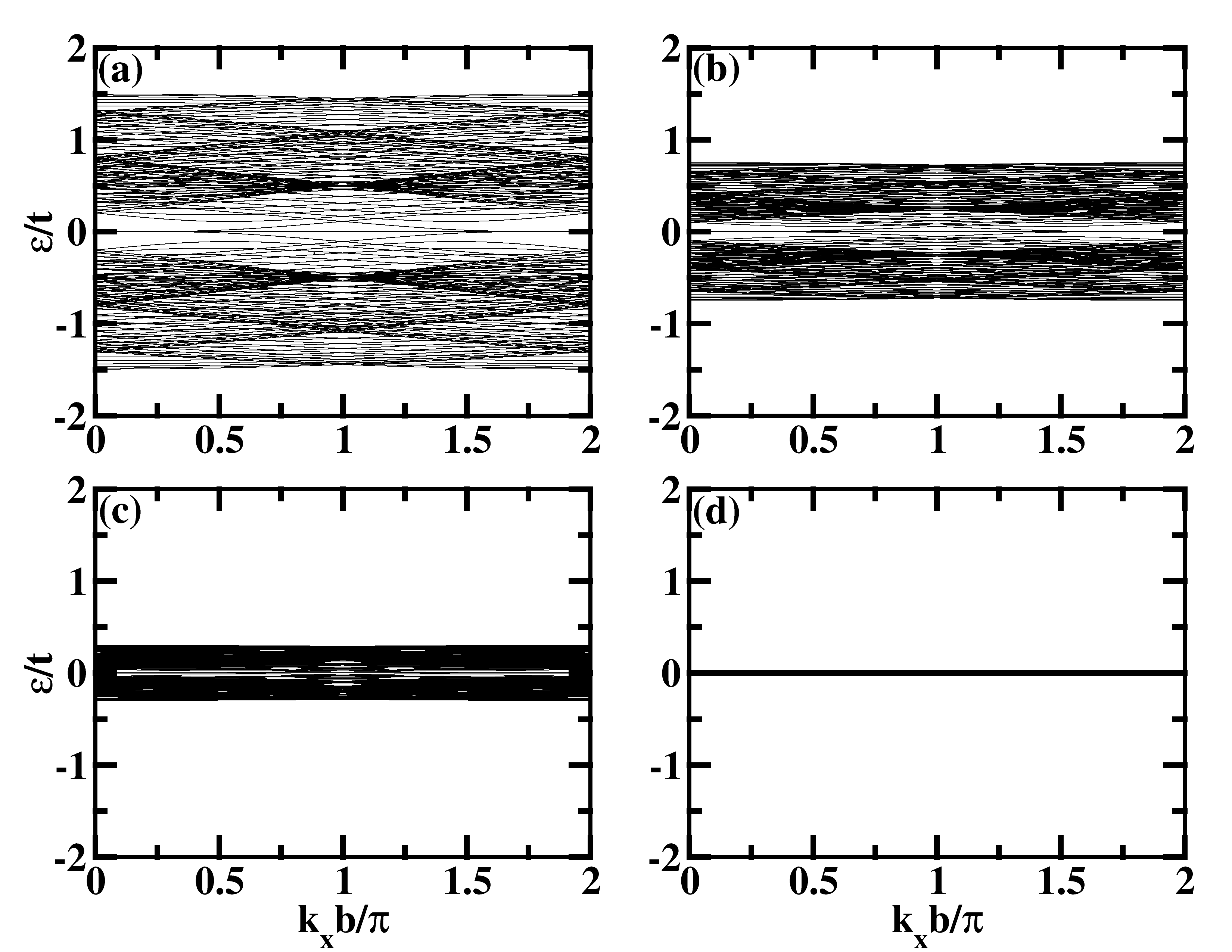}
    \caption{ZGNR irradiated with CPL in high-frequency regime with (a)$A_0 = 1.521$, (b)$A_0 = 1.954$, (c)$A_0 = 2.218$ and (d)$A_0 = 2.385$.}
    \label{fig:CPL_normal_very_high_omega}
\end{figure}
The renormalized hopping energy along all three nearest neighbours decreases equally with increasing amplitude. This decrease in hopping strength induces localization of electron along all three bonds and starts showing the properties of isolated atoms. 
\begin{equation}
\begin{split}
t_1^F &= t J_0(A_0a)\\
t_2^F &= t J_0(A_0a)\\
t_3^F &= t J_0(A_0a)
\end{split}\label{cpl_hopping}
\end{equation}

Figure \ref{fig:CPL_normal_very_high_omega} illustrates the quasienergy spectra of a ZGNR when subjected to CPL with very high-frequency. In panel \ref{fig:CPL_normal_very_high_omega}(a), the scenario is presented for $A_0 = 1.521$, resulting in a hopping strength of $t/2$. For panels \ref{fig:CPL_normal_very_high_omega}(b) and \ref{fig:CPL_normal_very_high_omega}(c), the hopping strength is reduced to $t/4$ and $t/10$, respectively. In panel \ref{fig:CPL_normal_very_high_omega}(d), with $A_0 = 2.385$, the hopping strength becomes $t/100$. It is evident from Fig. \ref{fig:CPL_normal_very_high_omega} that the band localization phenomenon begins at $\varepsilon = 0$, indicating properties similar to those of an isolated atom.

\section{ZGNRs with varying $l$ irradiated by CPL in low frequency regime. }
 
\begin{figure}[ht]
	\centering
	\includegraphics[width=9cm,height=8cm]{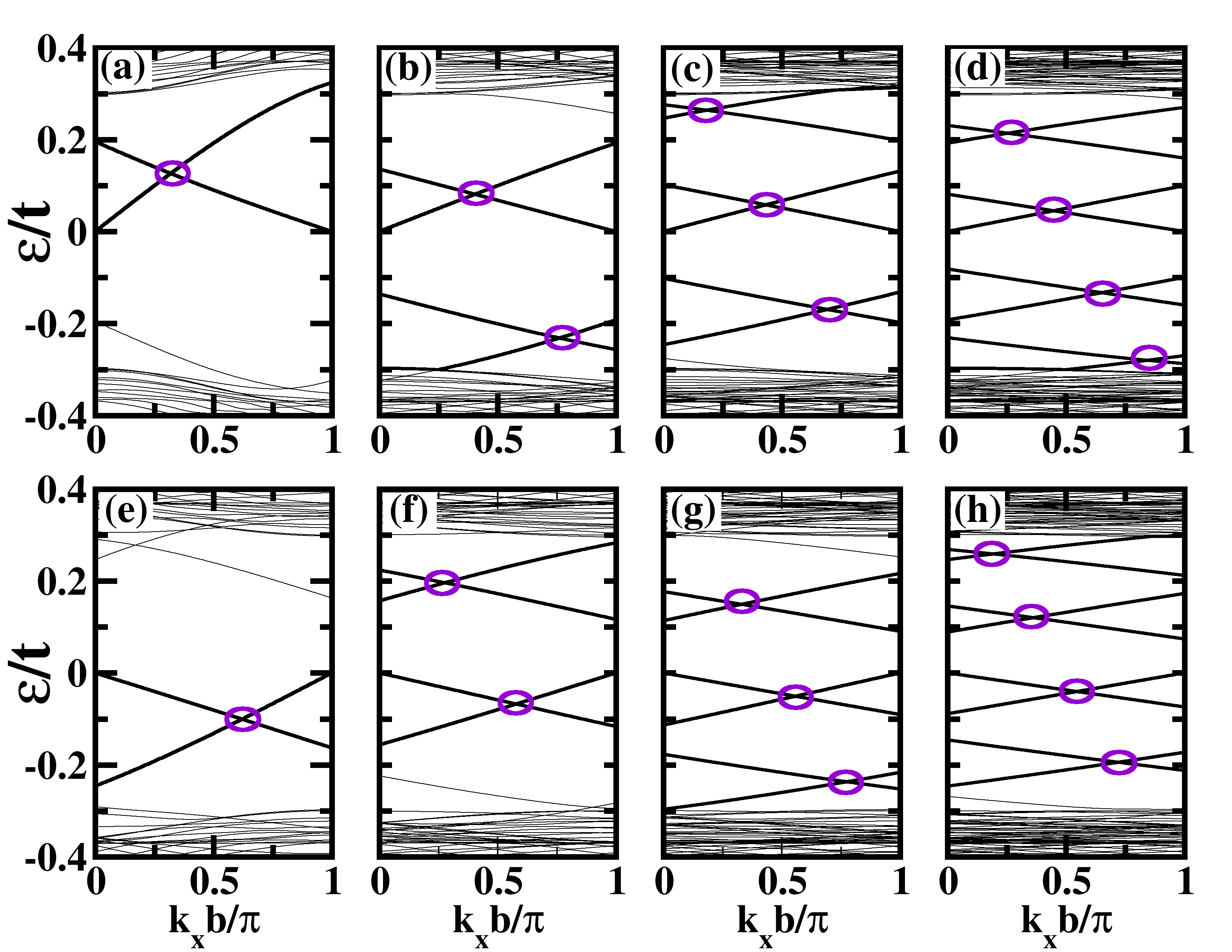}
	\caption{Quasienergy spectra of A-ZGNR for $A_0 =2.04$ and $\phi = -\pi/2$. (a) to (d) are for $l =2,4,6,8$ and (e) to (h) are for $l=3,5,7,9$. Encircled points represent the Dirac points. }
	\label{fig:merge_varying_l}
\end{figure}

As the separation between the periodic array of monovacancy defects increases, there is a corresponding increase in the parameter denoted as $l$. We have established a calibration of $l$ in relation to the number of Dirac points (denoted as $n$) that appear between the bulk valence and conduction bands, while keeping the field amplitude fixed at a value of $A_0 = 2.01$ and the phase at $\phi = -\pi/2$.
Figures \ref{fig:merge_varying_l}(a) to \ref{fig:merge_varying_l}(d) depict the quasienergy spectra for even values of $l$, specifically, 2, 4, 6, and 8. In parallel, Figs. \ref{fig:merge_varying_l}(e) to \ref{fig:merge_varying_l}(h) represent the quasienergy spectra for $l$ values of 3, 5, 7, and 9. A noticeable trend emerges: an increase in the value of $l$ leads to a corresponding increase in the value of $n$. This relationship arises because the multiple zone folding effect results in the crossing of multiple edge states in the system.
For situations where $l$ is an even number, the value of $n$ is simply half of $l$ (i.e. $n = l/2$). Conversely, when $l$ is an odd number, $n$ can be expressed as $l-1/2$. Therefore, by examining the quasienergy spectra and counting the number of Dirac points, one can readily determine the values of $l$.

\bibliography{main.bib}

\end{document}